\documentclass[conference]{IEEEtran}
\usepackage{stmaryrd}
\usepackage{amsfonts}
\usepackage[dvips]{graphicx}
\usepackage[latin1]{inputenc}
\usepackage{graphicx,times,amsmath,epsfig,amssymb,array,enumerate,algorithm, algorithmic} 

\newcommand{\bx}{\mathbf{x}}
\newcommand{\bw}{\mathbf{w}}
\newcommand{\bz}{\mathbf{z}}
\newcommand{\bT}{\mathbf{T}}

\newcommand{\bW}{\boldsymbol{W}}
\newcommand{\bZ}{\mathbf{Z}}
\newcommand{\bsx}{\boldsymbol{x}}
\newcommand{\bsw}{\boldsymbol{w}}
\newcommand{\bsr}{\boldsymbol{r}}
\newcommand{\bsv}{\boldsymbol{v}}
\newcommand{\bst}{\boldsymbol{t}}
\newcommand{\btheta}{\boldsymbol{\theta}}
\newcommand{\bgamma}{\boldsymbol{\gamma}}
\newcommand{\bbeta}{\boldsymbol{\beta}}
\newcommand{\bpsi}{\boldsymbol{\psi}}
\newcommand{\IR}{I\!\!R}

\IEEEoverridecommandlockouts    

\begin{document}

\title{\ \\ \LARGE\bf A regression model with a hidden logistic process for feature extraction from time series}

\author{Faicel Chamroukhi, Allou Sam\'{e}, G\'{e}rard Govaert and Patrice Aknin}

\maketitle

\begin{abstract}
A new approach for feature extraction from time series is proposed in this paper. This approach consists of a specific regression model incorporating a discrete hidden logistic process. The model parameters are estimated by the maximum likelihood method performed by a dedicated Expectation Maximization (EM) algorithm. The parameters of the hidden logistic process, in the inner loop of the EM algorithm, are estimated using a multi-class Iterative Reweighted Least-Squares (IRLS) algorithm. A piecewise regression algorithm and its iterative variant have also been considered for comparisons. An experimental study using simulated and real data reveals good performances of the proposed approach.
\end{abstract}


\section{Introduction}
\begin{figure}[!h]
  \centerline{\includegraphics[width=2.5in]{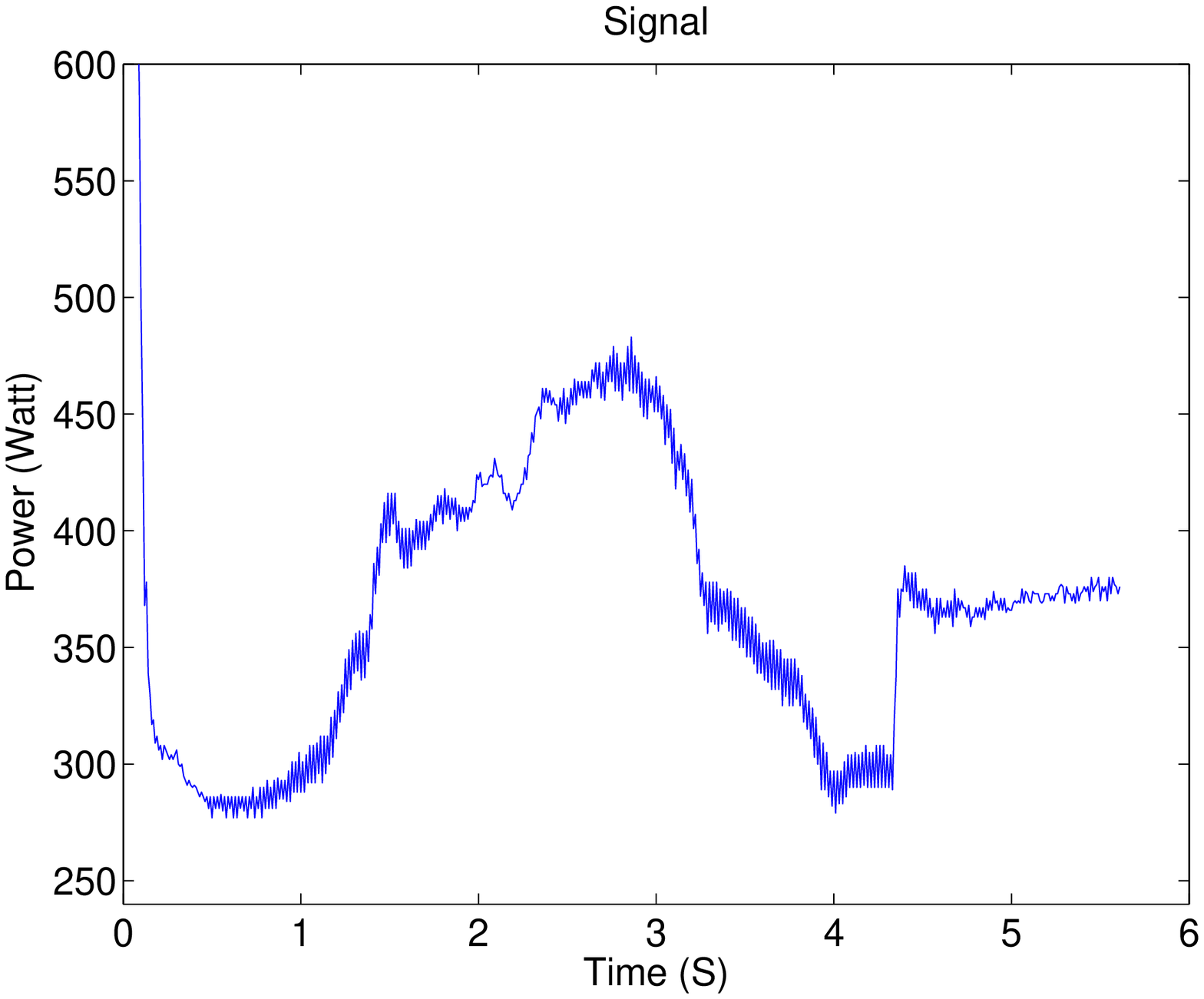}}
  \caption{Example of the electrical power consumed during a point operation.}
  \label{signal_intro}
 \end{figure}

\PARstart{I}{n} the context of the predictive maintenance of the french railway switches (or points) which enable trains to be guided from one track to another at a railway junction, we have been brought to extract features from switch operations signals representing the electrical power consumed during a point operation (see Fig. \ref{signal_intro}). The final objective is to exploit these parameters for the identification of incipient faults.

The switch operations signals can be seen as time series presenting non-linearities and various changes in regime. Basic linear regression can not be adopted for this type of signals because a constant linear relationship is not adapted. As alternative to linear regression, some authors use approaches based on a piecewise regression model \cite{McGee}\cite{brailovsky}\cite{ferrari1}. Piecewise regression is a segmentation method providing a partition of the data into $K$ segments, each segment being characterized by its mean curve (constant, polynomial, ...) and its variance in the Gaussian case. Under this type of modeling, the parameters estimation is generally based on a global optimization using dynamic programming \cite{bellman} like Fisher's algorithm \cite{fisher}. This algorithm optimizes an additive criterion representing a cost function over all the segments of the signal \cite{yveslechevalier90}\cite{brailovsky}. However, the dynamic programming procedure is known to be computationally expensive. An iterative algorithm can be derived to improve the running time of Fisher's algorithm as in \cite{sameSFC2007}. This iterative approach is a local optimization approach estimating simultaneously the regression model parameters and the transition points. These two approaches will be recalled in our work, where the second one will be extended to supposing different variances for the various segments instead of using a constant variance for all the segments. Other alternative approaches are based on Hidden Markov Models \cite{rabiner} in a context of regression \cite{fridman} where the model parameters are estimated by the Baum-Welch algorithm \cite{baum welch}.

The method we propose for feature extraction is based on a specific regression model incorporating a discrete hidden process allowing for abrupt or smooth transitions between various regression models. This approach has a connection with the switching regression model introduced by Quandt and Ramsey \cite{quandt} and is very linked to the Mixture of Experts (ME) model introduced by Jordan and Jacobs \cite{jordan HME} by the using of a time-dependent logistic transition function. The ME model, as discussed in \cite{waterhouse}, uses a conditional mixture modeling where the model parameters are estimated by the Expectation Maximization (EM) algorithm \cite{dlr}\cite{mclachlan EM}. 

This paper is organized as follows. Section 2 recalls the piecewise regression model and two techniques of parameter estimation using a dynamic programming procedure: the method of global optimization of Fisher and its iterative variant. Section 3 introduces the proposed model and section 4 describes the parameters estimation via the EM algorithm. The fifth section is devoted to the experimental study using simulated and real data.

\section{Piecewise regression}
\label{sec: picewise regression}

Let $\bx=(x_1,\ldots,x_n)$ be $n$ real observations of a signal or a time serie where $x_i$ is observed at time $t_i$. The piecewise regression model supposes that the signal presents unknown transition points whose indexes can be denoted by $\bgamma = (\gamma_1,\ldots,\gamma_{K+1})$ with $\gamma_1=0$ and $\gamma_{K+1}=n$. This defines a partition $P_{n,K}$ of the time serie into $K$ segments of lengths $n_1,\ldots,n_K$ such that:
\begin{equation}
P_{n,K} = \{\bsx_1,\ldots,\bsx_K\},
\end{equation}
with $\bsx_k = \{x_i|i\in I_k\}$ and $I_k = ]\gamma_{k},\gamma_{k+1}]$.

Thus, the piecewise regression model generating the signal $\bx$ is defined as follows:
\begin{eqnarray}
\forall i=1,\ldots,n,\quad x_i=
\left \{ \begin{tabular}{l}
$\bbeta^T_1\bsr_{i} + \sigma_1 \epsilon_{i} \quad  \mbox{ if } i\in I_1$\\
$\bbeta^T_2\bsr_{i} + \sigma_2\epsilon_{i} \quad  \mbox{ if } i\in I_2$\\
$\vdots$\\
$\bbeta^T_K\bsr_{i} +\sigma_K\epsilon_{i} \quad  \mbox{ if } i\in I_K$
\end{tabular}\right.,
\label{eq.piecewise regression model}
\end{eqnarray}
where $\bbeta_k$, is the $(p+1)$-dimensional coefficients vector of a $p$ degree polynomial associated to the $k^{th}$ segment, $k \in \{1,\ldots,K\}$, $\bsr_{i}=(1,t_i,\ldots,(t_i)^p)^T$ is the time dependent $(p+1)$-dimensional covariate vector associated to the parameter $\bbeta_{k}$ and the $\epsilon_{i}$ are independent random variables distributed according to a Gaussian distribution with zero mean and unit variance representing an additive noise on each segment $k$.

\subsection{Parameter estimation}
Under this model, the parameters estimation is performed by maximum likelihood. We assume 
a conditional independence of the data between the segments, and the data within a segment are also supposed to be conditionally independent. Thus, according to the model (\ref{eq.piecewise regression model}), the log-likelihood of the parameter vector $\bpsi=(\bbeta_1,\ldots,\bbeta_K,\sigma_1^2,\ldots,\sigma_K^2)$ and the transition points $\bgamma=(\gamma_{1},\ldots,\gamma_{K+1})$ characterizing the piecewise regression model is a sum of local log-likelihoods over all the segments and can be written as follows:
\begin{eqnarray}
L(\bpsi,\bgamma;\bx) &=& \sum_{k=1}^K \ell_k (\bbeta_k,\sigma_k^2;\bsx_k ),
\end{eqnarray}
where
\small{
\begin{eqnarray}
\ell_k (\bbeta_k,\sigma_k^2;\bsx_k ) &\!\!\!\!\!\!=\!\!\!\!\!\!&\log p(\bsx_k;\bbeta_k,\sigma_k^2)\nonumber \\
&\!\!\!\!\!\!=\!\!\!\!\!\!&\sum_{i\in I_k} \log \mathcal{N}(x_i;\bbeta_k^T\bsr_i,\sigma_k^2)\nonumber \\
&\!\!\!\!\!\!=\!\!\!\!\!\!&-\frac{1}{2} \sum_{i\in I_k} \Big[\log{\sigma_k^{2}} + \frac{(x_i-\bbeta_k^{T}\bsr_i )^2}{\sigma_k^2} \Big]+\mbox{c}_k,
\end{eqnarray}}
is the log-likelihood within the segment $k$ and $\mbox{c}_k$ is a constant. Thus, the log-likelihood 
is finally written as:
\begin{eqnarray}
\!\!\!\! L(\bpsi,\bgamma;\bx)&\!\!\!\!\!=\!\!\!\!\!&-\frac{1}{2}\sum_{k=1}^K \sum_{i\in I_k}\Big[\log{\sigma_k^{2}}+\frac{(x_i-\bbeta_k^{T}\bsr_i )^2}{\sigma_k^2}\Big] + \mbox{C},
\end{eqnarray}
where $\mbox{C}$ is a constant.

Maximizing this log-likelihood is equivalent to minimizing with respect to $\bpsi$ and $\bgamma$ the criterion
\begin{eqnarray}
J(\bpsi,\bgamma) &=&\sum_{k=1}^K \sum_{i\in I_k}\Big[\log{\sigma_k^{2}} + \frac{(x_i-\bbeta_k^{T}\bsr_i )^2}{\sigma_k^2}\Big] \nonumber\\
&=& \sum_{k=1}^K J_k(\bpsi,\gamma_k,\bgamma_{k+1}),
\label{eq.picewise_reg criterion}
\end{eqnarray}
 where $J_k(\bpsi,\gamma_k,\bgamma_{k+1}) \!=\!  \sum_{i=\gamma_k+1}^{\bgamma_{k+1}}\Big[\log{\sigma_k^{2}} + \frac{(x_i-\bbeta_k^{T}\bsr_i )^2}{\sigma_k^2}\Big]$.

\subsection{Fisher's algorithm for estimating the parameters of a piecewise regression model}

The optimization algorithm of Fisher is an algorithm based on dynamic programming, providing the optimal partition of the data by minimizing an additive criterion \cite{fisher}\cite{brailovsky}\cite{yveslechevalier90}. This algorithm minimizes the criterion $J$ or equivalently minimizes, with respect to $\bgamma$, the criterion
\begin{eqnarray}
C_K(\bgamma) &=& \min \limits_{\substack {\bpsi}} J(\bpsi,\bgamma)  \nonumber \\
&=& \sum_{k=1}^K \min \limits_{\substack {\bbeta_k,\sigma_k^2}} \sum_{i=\gamma_k+1}^{\gamma_{k+1}}\Big[\log{{\sigma}_k^2} +\frac{(x_i-{\bbeta}_k^T\bsr_i)^2}{{\sigma}_k^2}\Big],\nonumber \\
& = & \sum_{k=1}^K c(\gamma_k,\gamma_{k+1}),
\label{dynaic programming criterion}
\end{eqnarray}
with $c(\gamma_k ,\gamma_{k+1})= \sum_{i=\gamma_k+1}^{\gamma_{k+1}}\Big[\log{{\hat{\sigma}}_k^2} + \frac{(x_i-{\hat{\bbeta}}_k^T\bsr_i)^2}{{\hat{\sigma}^2}_k}\Big]$, where
\begin{eqnarray}
{\hat{\bbeta}}_k^T &=& \arg \min \limits_{\substack{\bbeta_k}} \sum_{i=\gamma_k+1}^{\gamma_{k+1}}(x_i-\bbeta_k^{T}\bsr_i)^2 \nonumber \\
&=& (\Phi_k^T\Phi_k)^{-1}\Phi_k^T\bsx_k,
\label{estimation beta}
\end{eqnarray}
$\Phi_k=[\bsr_{\gamma_k+1},\ldots,\bsr_{\gamma_{k+1}}]^T$ being the regression matrix associated to $\bsx_k$, and
\begin{eqnarray}
\hat{\sigma}_k^2 
&=& \frac{1}{ n _k} \sum_{i=\gamma_k+1}^{\gamma_{k+1}} (x_i-{\hat{\bbeta}}_k^T \bsr_i)^2,
\label{estimation sigma}
\end{eqnarray}
$n_k$ being the number of points of the segment $k$.

It can be observed that the criterion $C_K(\bgamma)$ is a sum of cost $c(\gamma_k,\gamma_{k+1})$ over the $K$ segments. Therefore, due to the additivity of this criterion, its optimization can be performed using a dynamic programming procedure \cite{yveslechevalier90}\cite{bellman}. Dynamic programming considers that an optimal partition of the data into $K$ segments is the union of an optimal partition into $K-1$ segments and a set of one segment. By introducing the cost
\begin{equation}
C_k(a,b) =  \sum_{\ell=1}^{k} \min \limits_{\substack {(\bbeta,\sigma^2)}} \sum_{i=a+1}^{b} \Big[\log{\sigma_k^{2}} +\frac{(x_i-\bbeta_k^{T}\bsr_i )^2}{\sigma_k^2}\Big],
\end{equation}
with $0\leq a <b \leq n$ and $k=1,\ldots,K$, the dynamic programming optimization algorithm runs as follows:

\subsubsection{Step 1. (Initialization)} This step consists of computing the cost matrix $C_1(a,b)$ corresponding to one segment $]a,b]$ for $0\leq a <b \leq n$. This cost matrix is computed as follows:
\begin{eqnarray}
C_1(a,b) 
&=&\min \limits_{\substack {(\bbeta,\sigma^2)}} \sum_{i=a+1}^{b} \Big[\log{\sigma^{2}} +\frac{(x_i-\bbeta^{T}\bsr_i)^2}{\sigma^2}\Big] \nonumber\\
&=& \sum_{i=a+1}^{b} \Big[\log{\hat{\sigma}^{2}} +\frac{(x_i-\hat{\bbeta}^{T}\bsr_i)^2}{\hat{\sigma}^2}\Big],
\label{dynamic programming criterion}
\end{eqnarray}
where $\hat{\bbeta}^{T}$ and $\hat{\sigma}^2$ are computed respectively according to the equations (\ref{estimation beta}) and (\ref{estimation sigma}) by replacing $]\gamma_k,\gamma_{k+1}]$ by $]a,b]\cdot$

\subsubsection{Step 2. (Dynamic programming procedure)} 
This step consists of computing the optimal cost $C_k(a,b)$ for $k=2,\ldots,K$ and $0\leq a <b \leq n$ using the following formula:
\begin{equation}
C_k(a,b) = \min \limits_{\substack {a\leq h\leq b}} \Big[C_{k-1}(a,h ) + C_1(h+1,b)\Big].
\end{equation}

\subsubsection{Step 3. (Finding the optimal partition)}
From the optimal costs $C_k(a,b)$, the optimal partition can be deduced (for more details see appendix A in \cite{brailovsky}).

While the Fisher algorithm provides the global optimum, it is known to be computationally expensive. To accelerate the convergence of this algorithm, one can derive an iterative variant as in \cite{sameSFC2007}.
\subsection{Iterative version of Fisher's algorithm}
In the iterative procedure, the criterion  $J(\bpsi,\bgamma)$ given by equation (\ref{eq.picewise_reg criterion}) is iteratively minimized  by starting from an initial value of the transition points $\bgamma^{(0)} = (\gamma_1^{(0)},\ldots,\gamma_{K+1}^{(0)})$  and alternating the two following steps until convergence:

\subsubsection{Regression step (at iteration $m$)}

Compute the regression model parameters $\bpsi^{(m)} = \{\bbeta_k^{(m)},\sigma_k^{2(m)}; k=1\ldots,K\}$ for the current values of the transition points $\bgamma^{(m)}$ by minimizing the criterion $J(\bpsi,\bgamma^{(m)})$ given by equation (\ref{eq.picewise_reg criterion}) with respect to $\bpsi$. This minimization consists of performing $K$ separated polynomial regressions and provides the following estimates:
\begin{eqnarray}
\bbeta_k^{T(m)} 
 &=& (\Phi_k^{T(m)}\Phi_k^{(m)})^{-1}\Phi_k^{T(m)}\bsx_k^{(m)},
\end{eqnarray}
where $\Phi_k^{(m)} = [\bsr_{\gamma_{k}^{(m)}+1},\ldots,\bsr_{\gamma_{k+1}^{(m)}}]^T$ is the regression matrix associated to the elements of the $k^{th}$ segment \linebreak $\bsx_k^{(m)}=\{x_i|i\in ]\gamma_{k}^{(m)},\gamma_{k+1}^{(m)}]\}$ at the iteration $m$,
\begin{eqnarray}
\sigma_k^{2(m)} 
&=& \frac{1}{n_k^{(m)}}\sum_{i=\gamma_{k}^{(m)}+1}^{\gamma_{k+1}^{(m)}} (x_i-\hat{\bbeta}_k^{T(m)})^2.
\end{eqnarray}

\subsubsection{Segmentation step (at iteration $m$)}

Compute the transition points $\bgamma^{(m+1)} = (\gamma_1^{(m+1)},\ldots,\gamma_{K+1}^{(m+1)})$ by minimizing the criterion $J(\bpsi,\bgamma)$ for the current value of $\bpsi = \bpsi^{(m)}$, with respect to $\bgamma$. This minimization can be performed using a dynamic programming procedure since the criterion $J(\bpsi^{(m)},\bgamma)$ is additive. However, in contrast with the previous method, where the computation of  the cost matrix $C_1(a,b)$ requires the computation of the regression model parameter $\{\hat{\bbeta}_k,\hat{\sigma}_k^{2}; k=1,\ldots,K\}$ for $0\leq a <b \leq n$, this iterative procedure simply uses the cost matrix computed with the current values of the parameters $\{\bbeta_k^{T(m)},\sigma_k^{2(m)}; k=1\ldots,K\}$ which improves the running time of the algorithm.

The next section presents the proposed regression model with a hidden logistic process.

\section{Regression model with a hidden logistic process}
\label{sec: regression model}

\subsection{The global regression model}

We represent a signal by the random sequence $\bx = (x_1,...,x_n)$ of $n$ real observations, where $x_i$ is observed at time $t_i$.
This sample is assumed to be generated by the following regression model with a discrete hidden logistic process $\bz=(z_1,\ldots,z_n)$, where $z_i\in\{1,\ldots,K \}$:
\begin{eqnarray}
\forall i=1,\ldots,n,\quad  \left \{ \begin{tabular}{l}
$x_i= \bbeta^T_{z_i}\bsr_{i} + \sigma_{z_i}\epsilon_{i} $
\\$\epsilon_{i} \sim \mathcal{N}(0,1)$\end{tabular}\right..
\label{eq.regression model}
\end{eqnarray}
In this model, $\bbeta_{z_i}$ is the $(p+1)$-dimensional coefficients vector of a $p$ degree polynomial, $\bsr_{i}=(1,t_i,\ldots,(t_i)^p)^T$ is the time dependent $(p+1)$-dimensional covariate vector associated to the parameter $\bbeta_{z_i}$ and the $\epsilon_{i}$ are independent random variables distributed according to a Gaussian distribution with zero mean and unit variance. This model can be reformulated in a matrix form by
\begin{equation}
\bx = \sum^{K}_{k=1} \bZ_k (\bT \bbeta_{k} + \sigma_k \boldsymbol{\epsilon}),\label{ecriture_vectorielle_du_modele}
\end{equation}
where $\bZ_k$ is a diagonal matrix whose diagonal elements are $(z_{1k},\ldots,z_{nk})$ with $z_{ik}=1$ if $x_i$ is generated by the $k^{th}$ regression model and $0$ otherwise, $\bT =\Big[\bsr_1,\ldots,\bsr_n\Big]^T$
is the $[n\times (p+1)]$ matrix of covariates, 
and $ \boldsymbol{\epsilon} =(\epsilon_1,\ldots,\epsilon_n)^T$ is the noise vector distributed according to a zero mean multidimensional Gaussian density with identity covariance matrix.

\subsection{The hidden logistic process}
\label{ssec: process}

This section defines the probability distribution of the process $\bz=(z_1,\ldots,z_n)$ that allows the switching from one regression model to another.

The proposed hidden logistic process supposes that the variables $z_i$, given the vector $\bst=(t_1,\ldots,t_n)$, are generated independently according to the multinomial distribution {\small$\mathcal{M}(1,\pi_{i1}(\bw),\ldots,\pi_{iK}(\bw))$}, where
\begin{equation}
\pi_{ik}(\bw)= p(z_i=k;\bw)=\frac{\exp{(\bsw_k^T\bsv_{i})}}{\sum_{\ell=1}^K\exp{(\bsw_{\ell}^T \bsv_{i})}}\quad,
\label{multinomial logit}
\end{equation}
is the logistic transformation of a linear function of the time-dependent covariate $\bsv_i=(1,t_i,\ldots,(t_i)^q)^T$, $\bsw_{k}=(\bsw_{k0},\ldots,\bsw_{kq})^T$ is the $(q+1)$-dimensional coefficients vector associated to the covariate $\bsv_i$ and $\bw = (\bsw_1,\ldots,\bsw_K)$. Thus, given the vector $\bst=(t_1,\ldots,t_n)$, the distribution of $\bz$ can be written as:
\begin{equation}
p(\bz;\bw)=\prod_{i=1}^n \prod_{k=1}^K \left(\frac{\exp{(\bsw_{k}^T\bsv_{i})}}{\sum_{\ell=1}^K\exp{(\bsw_{\ell}^T \bsv_{i})}}\right)^{z_{ik}} \enspace ,
\label{eq.hidden logistic process}
\end{equation}
where $z_{ik} = 1$ if $z_i=k$ i.e when $x_i$ is generated by the $k^{th}$ regression model, and $0$ otherwise.

\begin{figure}[!h]
\begin{tabular}{cc}
\includegraphics[width=4cm]{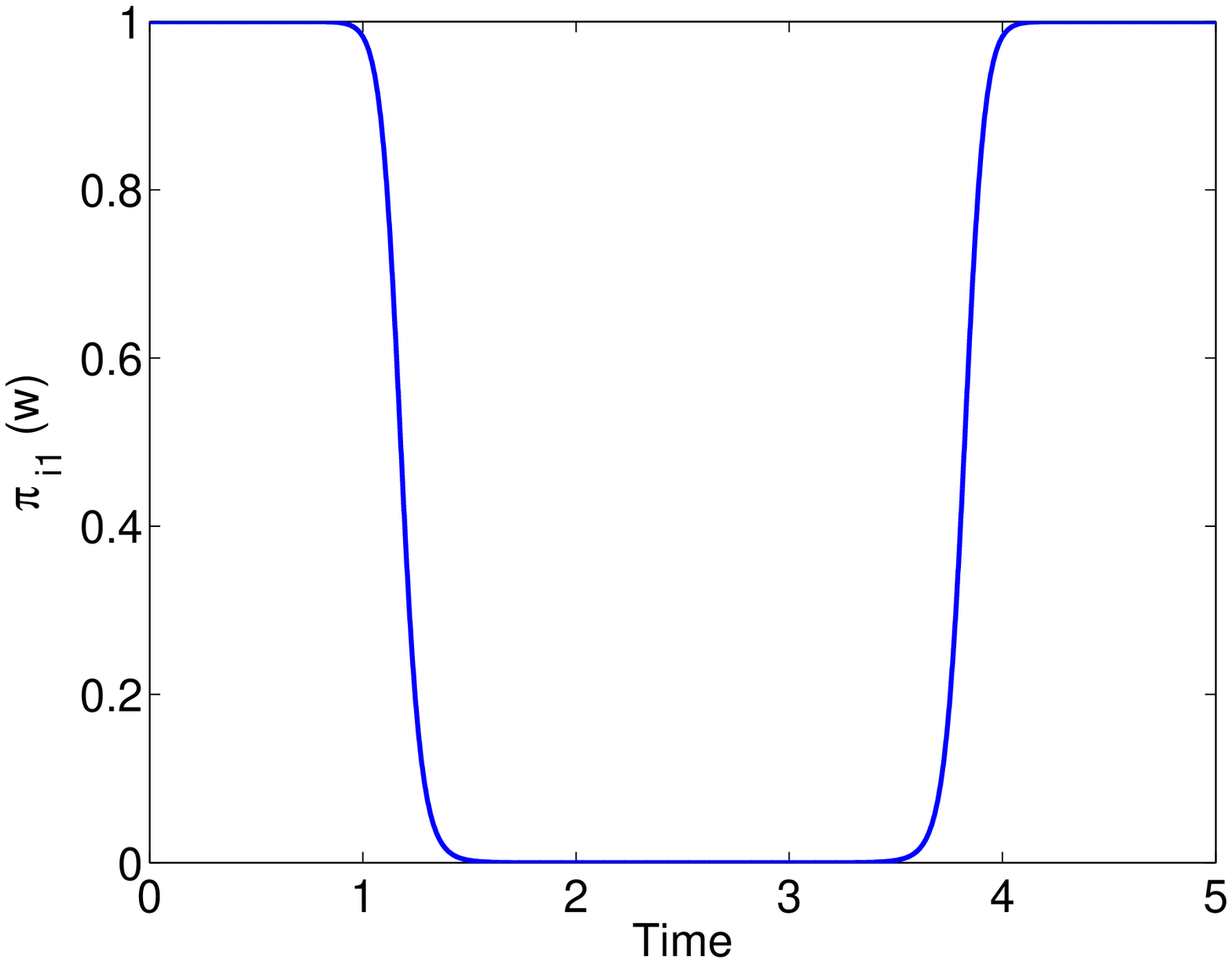}&
\includegraphics[width=4cm]{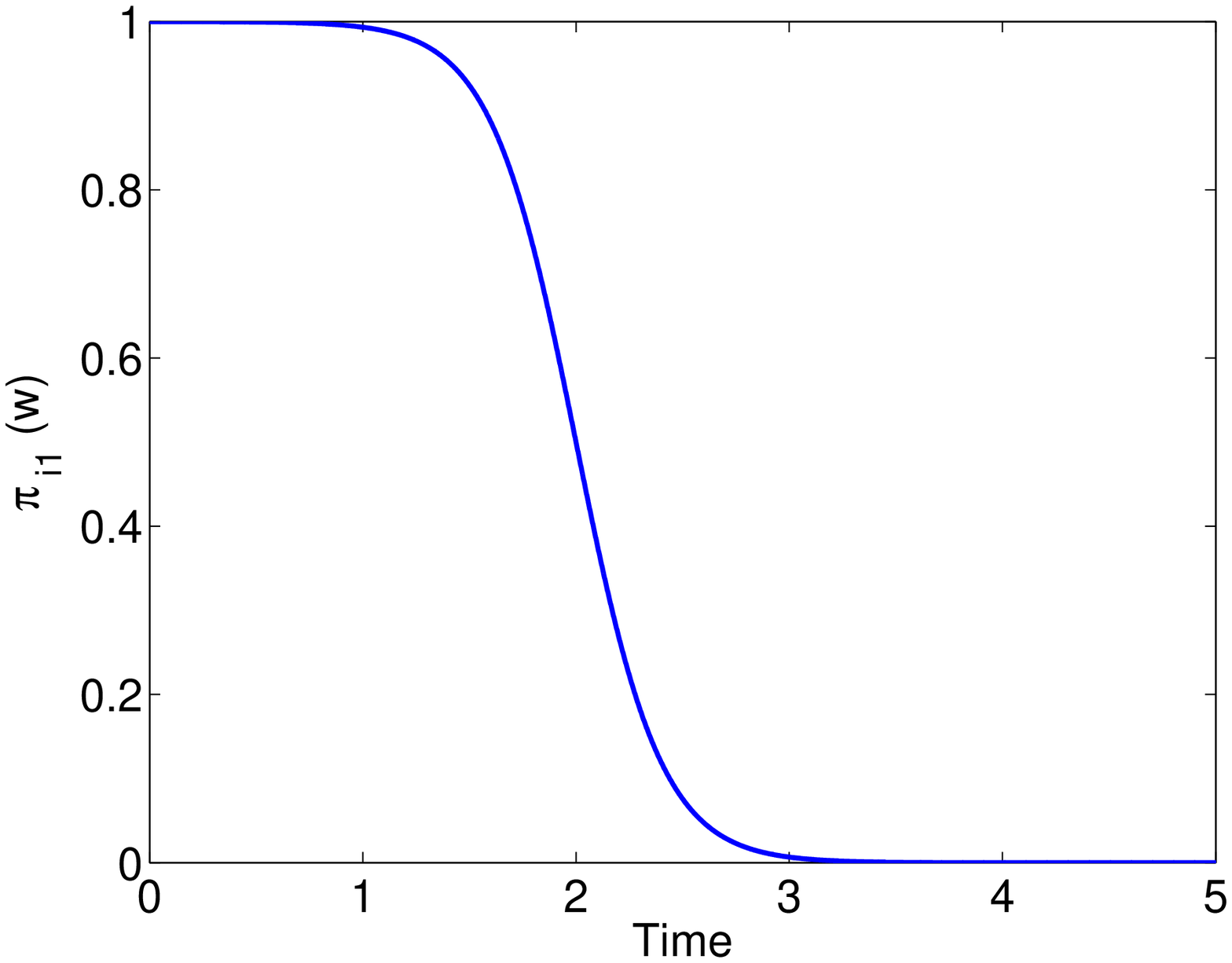}\\
(a) & (b)\\
\end{tabular}
\caption{Variation of $\pi_{i1}(\bw)$ over time for different values of the dimension $q$ of $\bsw_k$, with $K = 2$ and (a) $q= 2$ and $\bsw_1 = (-10,-20,-4)^T$, (b) $q=1$ and $\bsw_1 = (10,-5)^T$. For $q=0$, $\pi_{i1}(\bw)$ is constant over time.}
\label{logistic_function_k=2_q=012}
\end{figure}

The pertinence of the logistic transformation in terms of flexibility of transition can be illustrated through simple examples with $K=2$ components.
The first example is designed to show the effect of the dimension $q$ of $\bsw_k$ on the temporal variation of the probabilities $\pi_{ik}$. We consider 
 different values of the dimension $q$ ($q=0,1,2$) of $\bsw_k$. In that case, only the probability $\pi_{i1}(\bw)= \frac{exp(\bsw^T_1 \bsv_i)}{1+exp(\bsw^T_1 \bsv_i)}$ should be described, since $\pi_{i2}(\bw)=1-\pi_{i1}(\bw)$.
As shown in Fig. \ref{logistic_function_k=2_q=012}, the dimension $q$ controls the number of changes in the temporal variations of $\pi_{ik} $. In fact, the larger the dimension of $\bsw_k$, the more complex the temporal variation of $\pi_{ik}$.
More particularly, if the goal is to segment the signals into convex segments, the dimension $q$ of $\bsw_k$ must be set to $1$.

\begin{figure}[!h]
\begin{tabular}{cc}
\includegraphics[width=4cm]{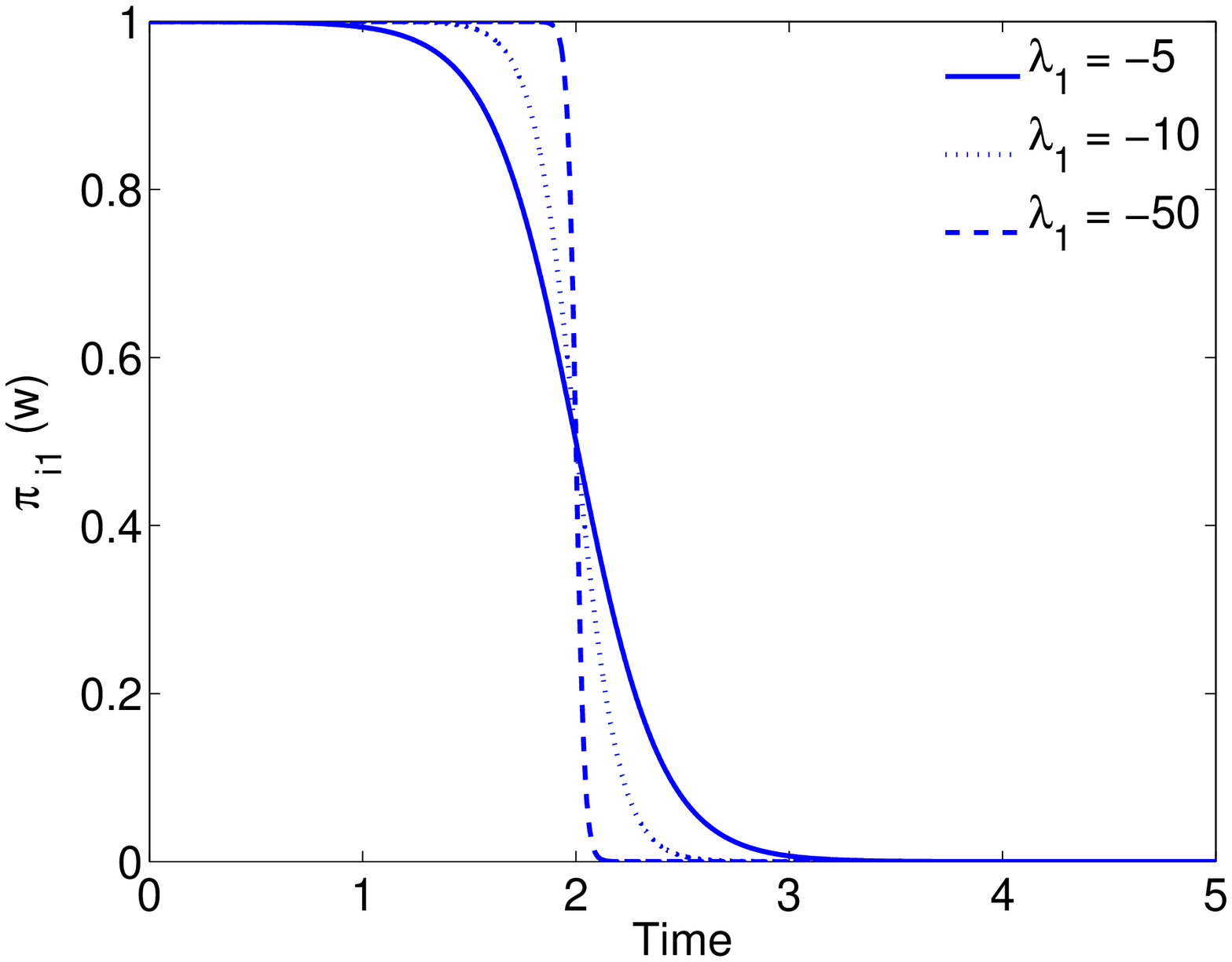} &
\includegraphics[width=4cm]{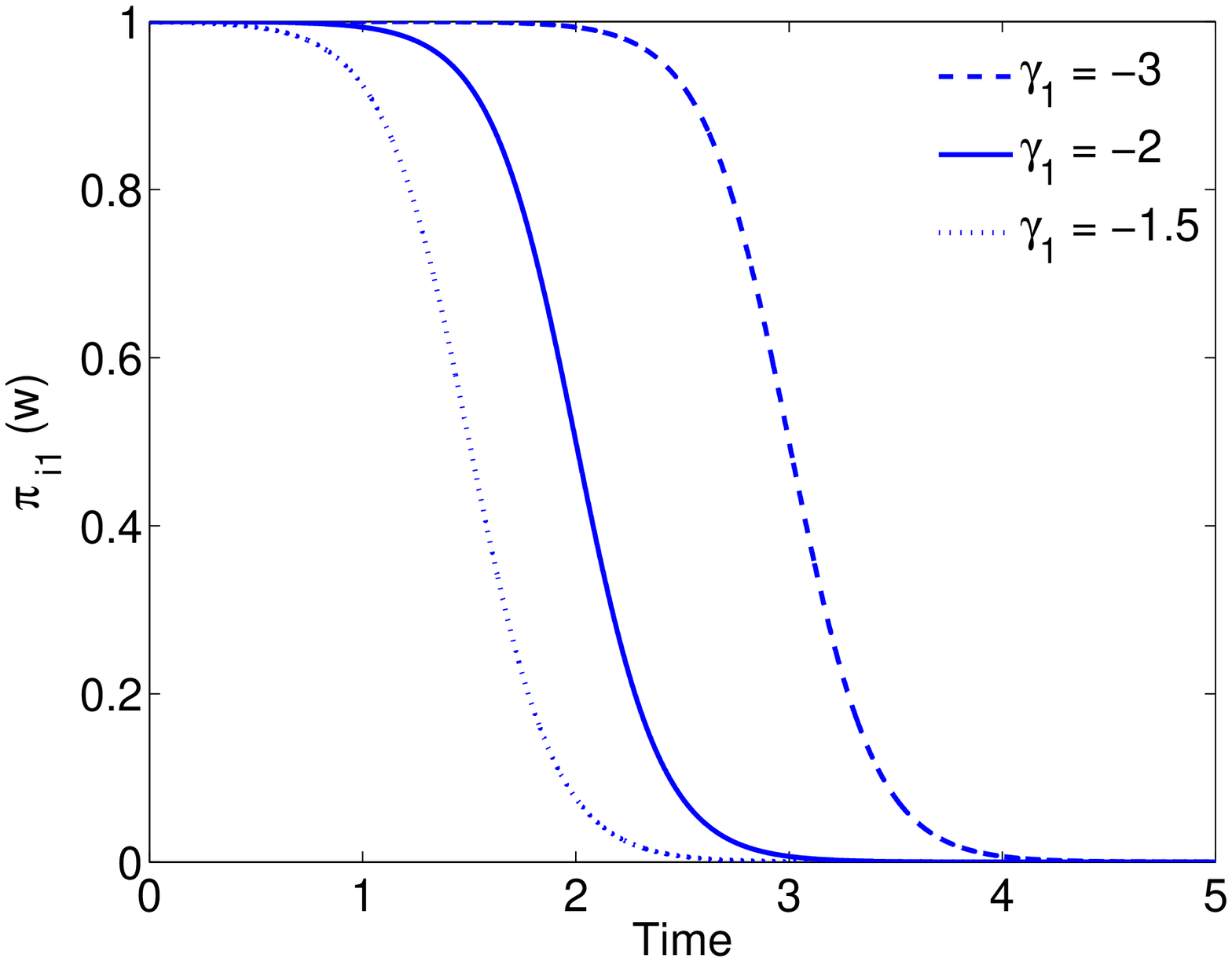}\\
(a) & (b)
\end{tabular}
\caption{Variation of $\pi_{i1}(\bw)$ over time for a dimension $q=1$ of $\bsw_k$ and (a) different values of $\lambda_k = \bsw_{k1}$ and (b) different values of $\gamma_k = \frac{\bsw_{k0}}{\bsw_{k1}}$.}
\label{logistic_function_k=2_q=1}
\end{figure}

For a fixed dimension $q$ of the parameter $\bsw_k$, the variation of the proportions $\pi_{ik}(\bw)$ over time, in relation to the parameter  $\bsw_k$, is illustrated by an example of 2 classes with $q=1$. For this purpose, we use the parametrization $\bsw_k =\lambda_k (\gamma_k, 1)^T$ of $\bsw_k$, where $\lambda_k= \bsw_{k1}$ and $\gamma_k = \frac{\bsw_{k0}}{\bsw_{k1}} \cdot$ As it can be shown in Fig. \ref{logistic_function_k=2_q=1} (a), the parameter $\lambda_k$ controls the quality of transitions between classes, more the absolute value of $\lambda_k$ is large, more the transition between the $z_i$ is abrupt, while the parameter $\gamma_k$ controls the transition time point by the means of the inflexion point of the curve (see Fig. \ref{logistic_function_k=2_q=1} (b)). In that case of 2 classes and $q=1$, the transition time point is the solution of $\bsw_{k0}+\bsw_{k1}t=0 $ which is $ t=-\gamma_k \cdot$

In this particular regression model, the variable $z_i$ controls the switching from a regression model to another one among $K$ regression models at each time $t_i$. Therefore, unlike basic polynomial regression models, which assume uniform regression parameters over time, the proposed model authorizes the polynomial coefficients to vary over time by switching from a regressive model to another.

\subsection{The generative model of signals}
	
The generative model of a signal from a fixed parameter \linebreak $\btheta=\{\bsw_k,\bbeta_k,\sigma^2_k ; k=1,\ldots,K\}$ consists in 2 steps: 
\begin{itemize}
\item generate the hidden process $(z_1,\ldots,z_n)$ with \linebreak $z_i \sim {\small \mathcal{M}(1,\pi_{i1}(\bw),\ldots,\pi_{iK}(\bw))}$,
\item  generate each observation $x_i$ according to the Gaussian distribution $\mathcal{N}(\cdot;\bbeta^{T}_{z_i}\bsr_i,\sigma^{2}_{z_i})$.
\end{itemize}

\section{Parameter estimation}
\label{sec: parameter estimation}

From the model 
(\ref{eq.regression model}), it can be proved that the random variable $x_{i}$ is distributed according to the normal mixture density
\begin{equation}
p(x_{i};\btheta)=\sum_{k=1}^K\pi_{ik}(\bw)\mathcal{N}\big(x_{i};\bbeta^T_k\bsr_{i},\sigma^2_k\big) \enspace,
\label{melange}
\end{equation}
where $\btheta=(\bsw_{1},\ldots,\bsw_{K},\bbeta_1,\ldots,\bbeta_K,\sigma^2_1,\ldots,\sigma^2_K)$
is the parameter vector to be estimated. The parameter $\btheta$ is estimated by the maximum likelihood method. As in the classic regression models we assume that, given $\bst =(t_1,\ldots,t_n)$,  the $\epsilon_i$ are independent. This also involves the independence of $x_i$ $(i=1,\ldots,n)$. The log-likelihood of $\btheta$ is then written as:
\begin{eqnarray}
L(\btheta;\bx)&\!\! = \!\!&\log \prod_{i=1}^np(x_i;\btheta)\nonumber\\
&\!\! = \!\!&\sum_{i=1}^{n}\log\sum_{k=1}^K \pi_{ik}(\bw)\mathcal{N}\big(x_{i};\bbeta^T_k\bsr_i,\sigma^2_k\big)\cdot
\end{eqnarray}
Since the direct maximization of this likelihood is not straightforward, we use the Expectation Maximization (EM) algorithm \cite{dlr}\cite{mclachlan EM} to perform the maximization.

\subsection{The dedicated EM algorithm}
\label{ssec. EM algortihm}

The proposed EM algorithm starts from an initial parameter $\btheta^{(0)}$ and alternates the two following steps until convergence:

\subsubsection{\textbf{E Step (Expectation)}}

This step consists of computing the expectation of the complete log-likelihood $\log p(\bx,\bz;\btheta)$, given the observations and the current value $\btheta^{(m)}$ of the parameter $\btheta$ ($m$ being the current iteration):
\begin{eqnarray}
\!\! \! \! \! Q(\btheta,\btheta^{(m)})&\!\!\!\!\!\! =\!\!\!\! \!\!& E\Big[\log p(\bx,\bz;\btheta)|\bx;\btheta^{(m)}\Big]\nonumber\\
&\!\!\!\!\!\! =\!\!\!\!\!\! &\sum_{i=1}^{n}\sum_{k=1}^K \tau^{(m)}_{ik}\log \Big[\pi_{ik}(\bw)\mathcal{N} \big(x_{i};\bbeta^T_k\bsr_{i},\sigma^2_k \big)\Big],
\end{eqnarray}
where
\begin{eqnarray}
 \tau^{(m)}_{ik} &\!\!=\!\!& p(z_{ik}=1|x_i;\btheta^{(m)}) \nonumber \\
 &\!\!=\!\!&\frac{\pi_{ik}(\bw^{(m)})\mathcal{N}(x_{i};\bbeta^{T(m)}_k\bsr_{i},\sigma^{2(m)}_k)}
{\sum_{\ell=1}^K\pi_{i \ell}(\bw^{(m)})\mathcal{N}(x_{i};\bbeta^{T(m)}_{\ell}\bsr_{i},\sigma^{2(m)}_{\ell})},
\label{eq.tik}
\end{eqnarray}

is the posterior probability that $x_i$ originates from the $k^{th}$ regression model.
\\As shown in the expression of $Q$, this step simply requires the computation of $\tau^{(m)}_{ik}$.

\subsubsection{\textbf{M step (Maximization)}}

In this step, the value of the parameter $\btheta$ is updated by computing the parameter $\btheta^{(m+1)}$ maximizing the conditional expectation $Q$
with respect to $\btheta$. The maximization of $Q$ can be performed by separately maximizing
\begin{equation}
Q_1(\bw)=\sum_{i=1}^{n}\sum_{k=1}^K \tau^{(m)}_{ik}\log \pi_{ik}(\bw)
\end{equation}
and, for all $k=1,\ldots,K$
\begin{eqnarray}
\!\! \! \! \!\!\! \! \! \!Q_2(\bbeta_k,\sigma^2_k) &\!\!\!\!=\!\!\!\!& \sum_{i=1}^{n} \tau^{(m)}_{ik}\log \mathcal{N}(x_{i};\bbeta^T_k\bsr_{i},\sigma^2_k) 
\end{eqnarray}
\\Maximizing  $Q_2$ with respect to the $\bbeta_k$ consists of analytically solving a weighted least-squares problem. The estimates are straightforward and are as follows:
\begin{eqnarray}
{\bbeta}_k^{T(m+1)} &=& \arg \min \limits_{\substack {\bbeta_k}} \sum_{i=1}^{n} \tau^{(m)}_{ik} ( x_i-\bbeta_k^{T}\bsr_i )^2 \nonumber \\
&=& (\bT^T\bW_k^{(m)}\bT)^{-1}\bT^T\bW_k^{(m)}\bx,
\label{estimation betaEM}
\end{eqnarray}
with $\bW_k^{(m)}$ is the $[n \times n]$ diagonal matrix of weights whose diagonal elements are $(\tau_{1k}^{(m)},\ldots,\tau_{nk}^{(m)})$ and $\bx = (x_1,\ldots,x_n)^T$ is the $[n\times 1]$-dimensional vector of observations.
    \\Maximizing  $Q_2$ with respect to the $\sigma_k^2$ provides the following estimates:
\begin{eqnarray}
\!\!{\sigma}_k^{2(m+1)} &\!\!\!\!=\!\!\!\!& \arg \min \limits_{\substack{\sigma_k^2}} \sum_{i=1}^{n} \tau^{(m)}_{ik}\Big[\log{\sigma_k^{2}} + \frac{(x_i-\bbeta_k^{T}\bsr_i )^2}{\sigma_k^2}\Big] \nonumber \\
&\!\!\!\!=\!\!\!\!& \frac{1}{\sum_{i=1}^{n} \tau^{(m)}_{ik}}\sum_{i=1}^{n} \tau^{(m)}_{ik} (x_i-{\bbeta}_k^{T(m+1)}\bsr_i)^2 \cdot
\label{estimation sigmaEM}
\end{eqnarray}
\\The maximization of $Q_1$ with respect to $\bw$ is a multinomial logistic regression problem weighted by the $\tau^{(m)}_{ik}$. We use a multi-class Iterative Reweighted Least Squares (IRLS) algorithm \cite{irls}\cite{krishnapuram}\cite{chen99} to solve it. The IRLS algorithm is detailed in the following section.

\subsubsection{The Iteratively Reweighted Least Squares (IRLS) algorithm}
\label{ssec: IRLS}
The IRLS algorithm is used to maximize $Q_1(\bw)$ 
with respect to the parameter $\bw$, in the M step, at each iteration $m$ of the EM algorithm. To estimate the parameters vector $\bw =(\bsw_1,\ldots,\bsw_{K})$, since $\sum_{k=1}^{K}\pi_{ik}(\bw)=1$, 
$\bsw_K$ is set to the null vector to avoid the identification problems.
The IRLS algorithm is equivalent to the Newton-Raphson algorithm, which consists of starting from a vector $\bw^{(0)}$, and updating the estimation of $\bw$ as follows:
\begin{equation}
\bw^{(c+1)}=\bw^{(c)}-\Big[{H(\bw^{(c)})}\Big]^{-1}g(\bw^{(c)})\enspace,
\label{eq.IRLS}
\end{equation}
where $ H(\bw^{(c)})$ and $g(\bw^{(c)})$ are respectively the Hessian and the gradient of $Q_1(\bw)$ evaluated at $\bw = \bw^{(c)}$. In \cite{krishnapuram}, authors use an approximation of the Hessian matrix to accelerate the convergence of the algorithm, while, in our case we use the exact Hessian matrix to perform well the maximum likelihood estimation as noticed in \cite{chen99}. Since there are $K-1$ parameters vectors $\bsw_1,\ldots,\bsw_{K-1}$ to be estimated, the Hessian matrix $H(\bw^{(c)})$ consists of $(K-1)\times(K-1)$ block matrices $H_{k\ell}(\bw^{(c)})(k,\ell=1,\ldots,K-1)$ \cite{chen99} where :
\begin{eqnarray}
 H_{k\ell}(\bw^{(c)})&=\!\!\!&\frac{\partial^2 Q_1(\bw)}{\partial \bsw_k \partial \bsw_{\ell}}\Big|_{\bw=\bw^{(c)}}\nonumber \\
		   &=\!\!\!&-\sum_{i=1}^{n} \pi_{ik}(\bw^{(c)}) [\delta_{k\ell} - \pi_{i \ell}(\bw^{(c)})] \bsv_i {\bsv_{i}}^T,
\end{eqnarray}
where $\delta_{k\ell}$ is the kronecker symbol ($\delta_{k\ell}$ = 1 if $k=\ell$, 0 otherwise).
The gradient of $Q_1(\bw)$ consists of $K-1$ gradients corresponding to the vectors $\bsw_k$ for $k=1,\ldots,K-1$ and is computed as follows:
\begin{eqnarray} 	
g(\bw^{(c)}) &\!\!=\!\!& \frac{\partial Q_1(\bw)}{\partial \bw}\Big|_{\bw=\bw^{(c)}}\nonumber\\
       &\!\!=\!\!& [g_{1}(\bw^{(c)}),\ldots,g_{K-1}(\bw^{(c)})]^{T}\enspace \!\!,
\end{eqnarray}
with
\begin{eqnarray} 	
\!\!\!\!\!\! g_{k}(\bw^{(c)}) &\!\!\!\!=\!\!\!\!& \frac{\partial Q_1(\bw)}{\partial \bsw_k}\Big|_{\bw=\bw^{(c)}}\nonumber\\
&\!\!\!\!=\!\!\!\!&\sum_{i=1}^{n} [\tau^{(m)}_{ik} - \pi_{ik}(\bw^{(c)})] \bsv^{T}_i \!\!\enspace;\enspace \!\! k=1,\ldots,K-1.
\end{eqnarray}
Applying algorithm (\ref{eq.IRLS}) provides the parameter $\bw^{(m+1)}$.

Algorithm ($1$) summarizes the proposed algorithm.
\begin{algorithm}
\caption{Pseudo code for the proposed algorithm.}
\begin{algorithmic}
\STATE \textbf{Initialize:}
\STATE fix a threshold $\epsilon>0$  \quad ;\quad  $m \leftarrow 0$ (iteration)
\STATE choose an initial $\btheta^{(m)}\!\!=\!\!\{\bsw_k^{(m)}, \bbeta_k^{(m)}, \sigma_k^{2(m)}; k\!\!=\!\!1,\ldots,K\}$
\item Compute the initial value of $\pi_{ik}^{(m)}$ for $i=1,\ldots,n$ and $k=1,\ldots,K$ using equation (\ref{multinomial logit})
\WHILE {increment in log-likelihood $> \epsilon$}
\STATE \{\textbf{E} step\}: Compute the $\tau_{ik}^{(m)}$ for $i=1,\ldots,n$ and $k=1,\ldots,K$ using equation (\ref{eq.tik})
\STATE \{\textbf{M} step\}: for $k=1,\ldots,K$
\STATE Compute $\bbeta_k^{(m+1)}$ using equation (\ref{estimation betaEM})
\STATE Compute $\sigma_k^{2(m+1)}$ using equation (\ref{estimation sigmaEM})
\STATE compute $\bw^{(m+1)}$ using the IRLS algorithm:
\STATE \{\textbf{IRLS} loop\}:
\STATE \textbf{Initialize:}
\STATE set a threshold $\delta>0$ \quad ;\quad $c\leftarrow 0$ (iteration)
\STATE set $\bw^{(c)} = \bw^{(m)}$
\WHILE{increment in $Q_1(\bw)>\delta$}
\STATE Compute $\pi_{ik}^{(c)}$ using equation (\ref{multinomial logit})
\STATE Compute $ \bw^{(c+1)}$ using  equation (\ref{eq.IRLS})
\STATE $c \leftarrow c+1$
\ENDWHILE
\STATE$\bw^{(m+1)} \leftarrow \bw^{(c)}$
\STATE$\pi_{ik}^{(m+1)} \leftarrow \pi_{ik}^{(c)}$ for $i=1,\ldots,n$ and $k=1,\ldots,K$
\STATE $m \leftarrow m+1$
\ENDWHILE
\STATE $\hat{\btheta}=(\bsw_1^{(m)},\ldots,\bsw_{K}^{(m)},\bbeta_1^{(m)},\ldots,\bbeta_K^{(m)},\sigma_1^{2(m)},\ldots,\sigma_K^{2(m)})$
\end{algorithmic}
\end{algorithm}

\subsection{Denoising and segmenting a signal}
\label{ssse: estimation of the denoised signal}

In addition to providing a signal parametrization, the proposed approach can be used to denoise and segment signals.
The denoised signal can be approximated by the expectation $E(\bx;\hat{\btheta}) = \big(E(x_1;\hat{\btheta}),\ldots,E(x_n;\hat{\btheta}) \big)$ where
\begin{eqnarray}
E(x_i;\hat{\btheta}) &=& \int_{\IR}x_i p(x_i;\hat{\btheta})dx_i\nonumber\\
			&=& \sum_{k=1}^{K} \pi_{ik}(\hat{\bw})\hat{\bbeta}^T_k \bsr_{i} \enspace , \forall i=1,\ldots,n,
\end{eqnarray}
and $\hat{\btheta} = (\hat{\bw},\hat{\bbeta}_1,\ldots,\hat{\bbeta}_K,\hat{\sigma}^2_1,\ldots,\hat{\sigma}^2_K)$ is the parameters vector obtained at the convergence of the algorithm. The matrix formulation of the approximated signal $\hat{\bx} = E(\bx;\btheta)$ is given by:
\begin{equation}
\hat{\bx} = \sum^{K}_{k=1} \hat{\mathbf{\Pi}}_{k} \bT \hat{\bbeta}_{k},
\label{eq. signal expectation}
\end{equation}
where $\hat{\mathbf{\Pi}}_{k}$ is a diagonal matrix whose diagonal elements are the proportions $(\pi_{1k}({\hat{\bw}}),\ldots,\pi_{nk}({\hat{\bw}}))$ associated to the $k^{th}$ regression model. On the other hand, a signal segmentation can also be deduced by computing the estimated label $\hat{z_i}$ of $x_i$ according to the following rule:
\begin{equation}
\hat{z_i} = \arg \max \limits_{\substack {1\leq k\leq K}} \pi_{ik}(\hat{\bw})\enspace, \quad  \forall i=1,\ldots ,n.
\label{partitionEMLogistique}
\end{equation}
\subsection{Model selection}

In a general use of the proposed model, the optimal values of $(K,p,q)$ can be computed by using the Bayesian Information Criterion \cite{BIC criterion} which is a penalized likelihood criterion, defined by
\begin{equation}
BIC(K,p,q) = L(\hat{\btheta};\bx) - \frac{\nu(K,p,q)\log(n)}{2}\enspace,
\end{equation}
where $\nu(K,p,p) = K(p+q+3)-(q+1)$ is the number of parameters of the model and $L(\hat{\btheta};\bx)$ is the log-likelihood obtained at the convergence of the EM algorithm. If the goal is to segment the data into convex segments $q$ must be set to $1$.
\section{Experiments}
\label{sec: experiments}

This section is devoted to the evaluation of the proposed algorithm using simulated and real data sets. For this purpose, the proposed approach is compared with the piecewise regression algorithm of Fisher and its iterative version. All the signals have been simulated from the piecewise regression model given by equation (\ref{eq.piecewise regression model}). Three evaluation criteria are used in the simulations.
\begin{itemize}
\item the first one is the misclassification rate between the simulated partition $P$ and the estimated partition $\hat{P}$,
\item the second one is the mean square error  between the expectations computed with the true parameter $\btheta$ and the estimated parameter $\hat{\btheta}$: $\frac{1}{n}\sum_{i=1}^{n}[E(x_i;\btheta)-E(x_i;\hat{\btheta})]^2$ where $E(x_i;\hat{\btheta})$ is computed according to equation (\ref{eq. signal expectation}) for the proposed model, and $E(x_i;\hat{\btheta}) = \bbeta_{\hat{z}_i}^T \bsr_i$ for the piecewise regression models. This error is used to asses the signal in terms of signal denoising and we call it the error of denoising.
\item the third criterion is the running time.
\end{itemize}

\subsection{Simulated signals}

\subsubsection{Protocol of simulations}

\begin{figure}[!h]
\centering
\includegraphics[width=4.25cm,height=3.8cm]{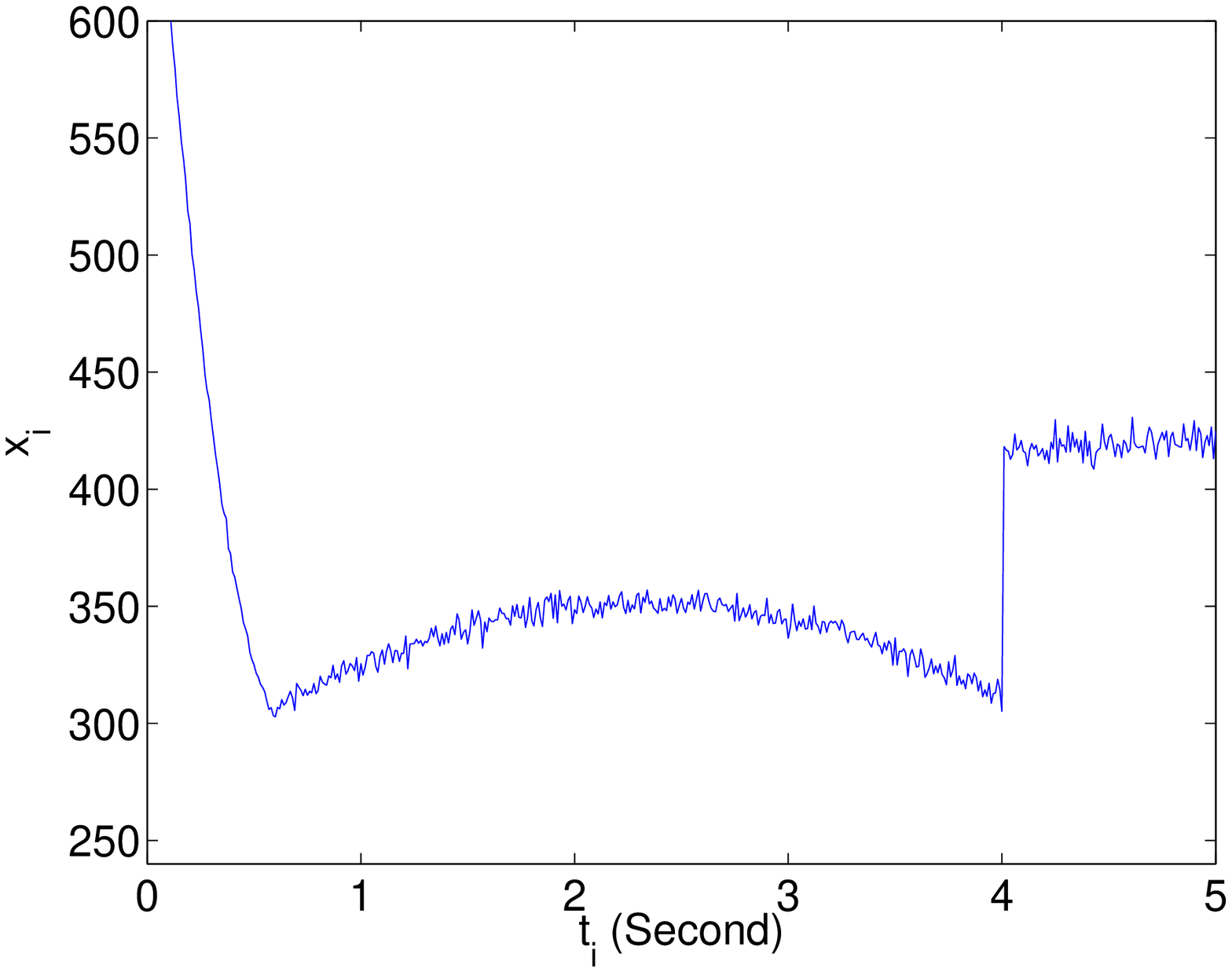}
\includegraphics[width=4.25cm,height=3.8cm]{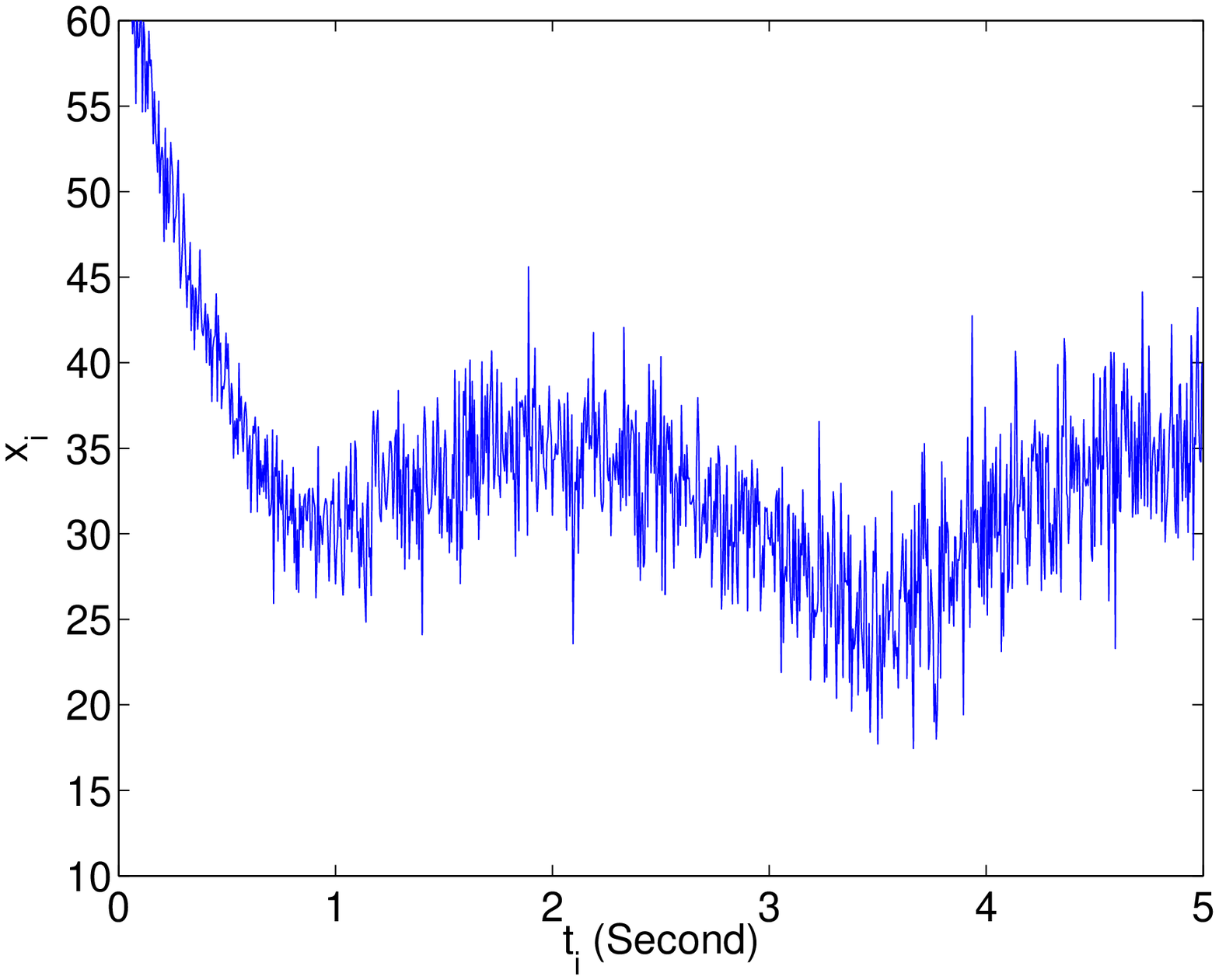}
(a)\hspace{3.5cm}(b)
\caption{Simulated signal for the first situation (a) and the second situation (b) for $n=1000$.}
\label{fig.example of simulated signals in picewise regression}
\end{figure}

For all the simulations, we set the number of segments (respectively the number of states of the hidden variable $z_i$ for the proposed model) to $K=3$ and the order of polynomial to $p=2$. We choose the value $q=1$ which guarantees a segmentation into contiguous intervals. We consider that all signals are observed over $5$ seconds (the time interval being fixed to $[0,5]$ Seconds) with a constant period of sampling $\Delta t=t_i-t_{i-1}$ depending on the sample size $n=100,200,...,1000$. For each size $n$ we generate 20 samples. The values of assessment criteria are averaged over the 20 samples. Two situations have been considered for simulations.
\begin{itemize}
\item situation1: the transition time points are set to $(0, 0.6, 4, 5)$ seconds, which correspond  $\gamma_1=0$, $\gamma_2=\frac{0.6}{\Delta t}$, $\gamma_3= \frac{4}{\Delta t}$ and $\gamma_4= \frac{5}{\Delta t}\cdot$ The set of parameters of simulations $\{\beta_k,\sigma_k^2; k=1,\ldots,K\}$ corresponding to this situation is given by table \ref{table. parameters of simulations sit1},
\item situation2: the transition time points are set to $(0, 1, 3.5, 5)$ seconds, which correspond to $\gamma_1=0$, $\gamma_2=\frac{1}{\Delta t}$, $\gamma_3= \frac{3.5}{\Delta t}$ and $\gamma_4= \frac{5}{\Delta t}\cdot$ The set of parameters of simulations $\{\beta_k,\sigma_k^2; k=1,\ldots,K\}$ corresponding to this situation is given by table \ref{table. parameters of simulations sit2}.
\end{itemize}
Fig. \ref{fig.example of simulated signals in picewise regression} shows an example of simulated signals for the two situations.
\begin{table}[htbp]
\centering
\begin{tabular}{lll}
\hline
$\bbeta_{1} = (735,-1320,1000)^T$ & $\sigma^{2}_{1} = 4$\\
$\bbeta_{2} = (270,60,-15)^T$ & $\sigma^{2}_{2} = 10$ \\
$\bbeta_{3} =(320,40,-4)^T$ & $\sigma^{2}_{3} = 15$\\
\hline
\end{tabular}
\caption{Parameters of simulations for situation 1.}
\label{table. parameters of simulations sit1}
\end{table}
\begin{table}[htbp]
\centering
\begin{tabular}{ll}
\hline
$\bbeta_{1} = (65,-70,35)^T$ & $\sigma^{2}_{1} = 4$\\
$\bbeta_{2} = (15,20,-5)^T$ & $\sigma^{2}_{2} = 10$ \\
$\bbeta_{3} =(-90,50,-5)^T$ & $\sigma^{2}_{3} = 15$\\
\hline
\end{tabular}
\caption{Parameters of simulations for situation 2.}
\label{table. parameters of simulations sit2}
\end{table}


\subsubsection{Strategy of initialization}
\label{ssec:initializations}

The proposed algorithm is initialized as follows:
\begin{itemize}
\item $\bsw_k=(0,\ldots,0)^T$ $\forall k=1,\ldots,K$,
\item to initialize $\bbeta_k$, we segment the signal uniformly into K segments and on each segment $k$ we fit a regression model, characterized by $\bbeta_k$,
\item $\sigma^2_k=1$ for $k=1,\ldots,K$.
\end{itemize}
For the iterative algorithm based on dynamic programming, several random initializations are used in addition to one initialization consisting of segmenting the signal into $K$ uniform segments, and the best solution corresponding to the smallest value of the criterion $J(\bpsi,\bgamma)$ is then selected. 
In the random initializations, the condition that the transition points are ordered in the time is respected. The algorithm is stopped when the increment in the criterion $J(\bpsi,\bgamma)$ is below $10^{-6}$.
\subsection{Results}
\begin{figure}[!h]
\centering
\begin{tabular}{c}
\includegraphics[width=7cm,height=6cm]{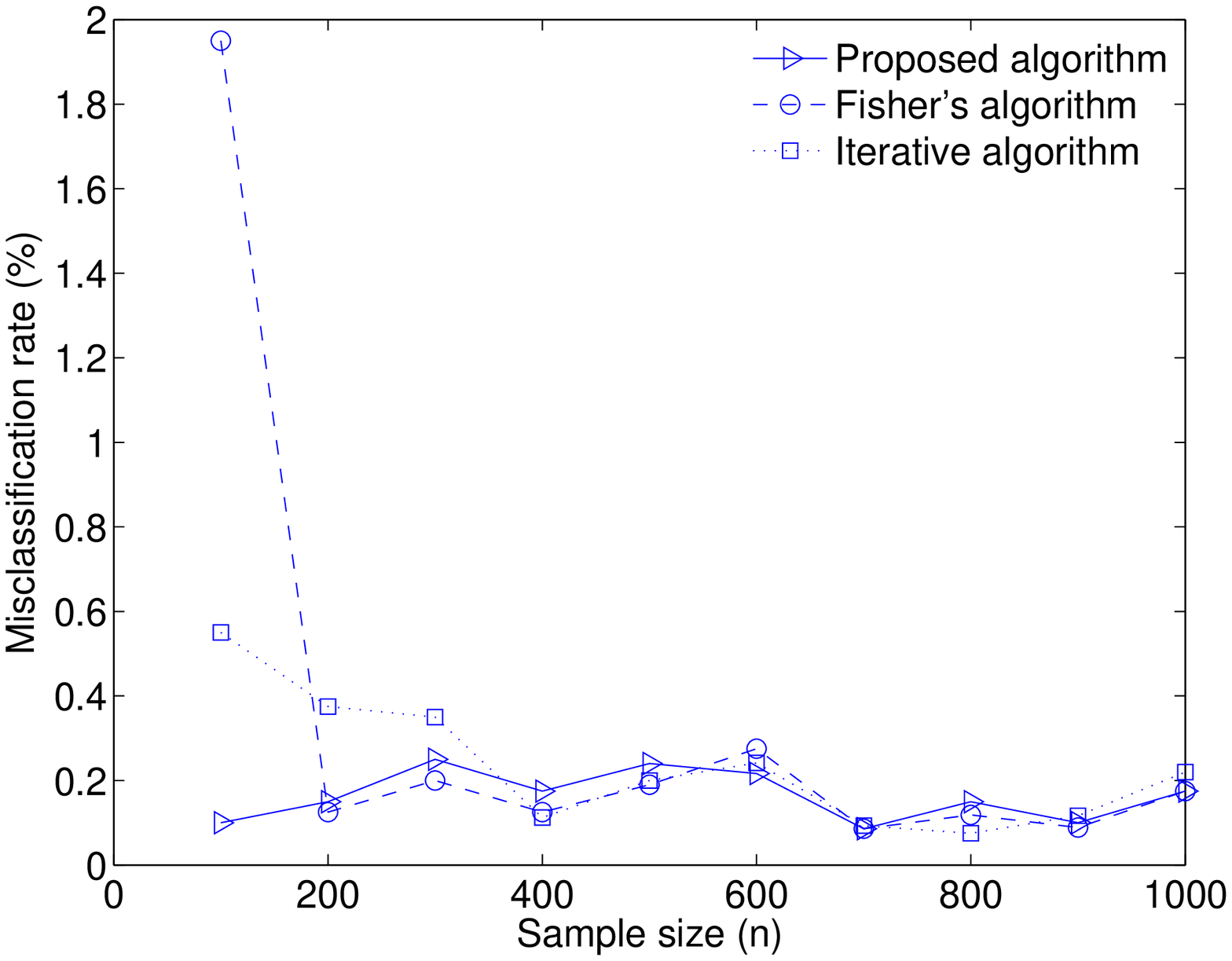}\\
\includegraphics[width=7cm,height=6cm]{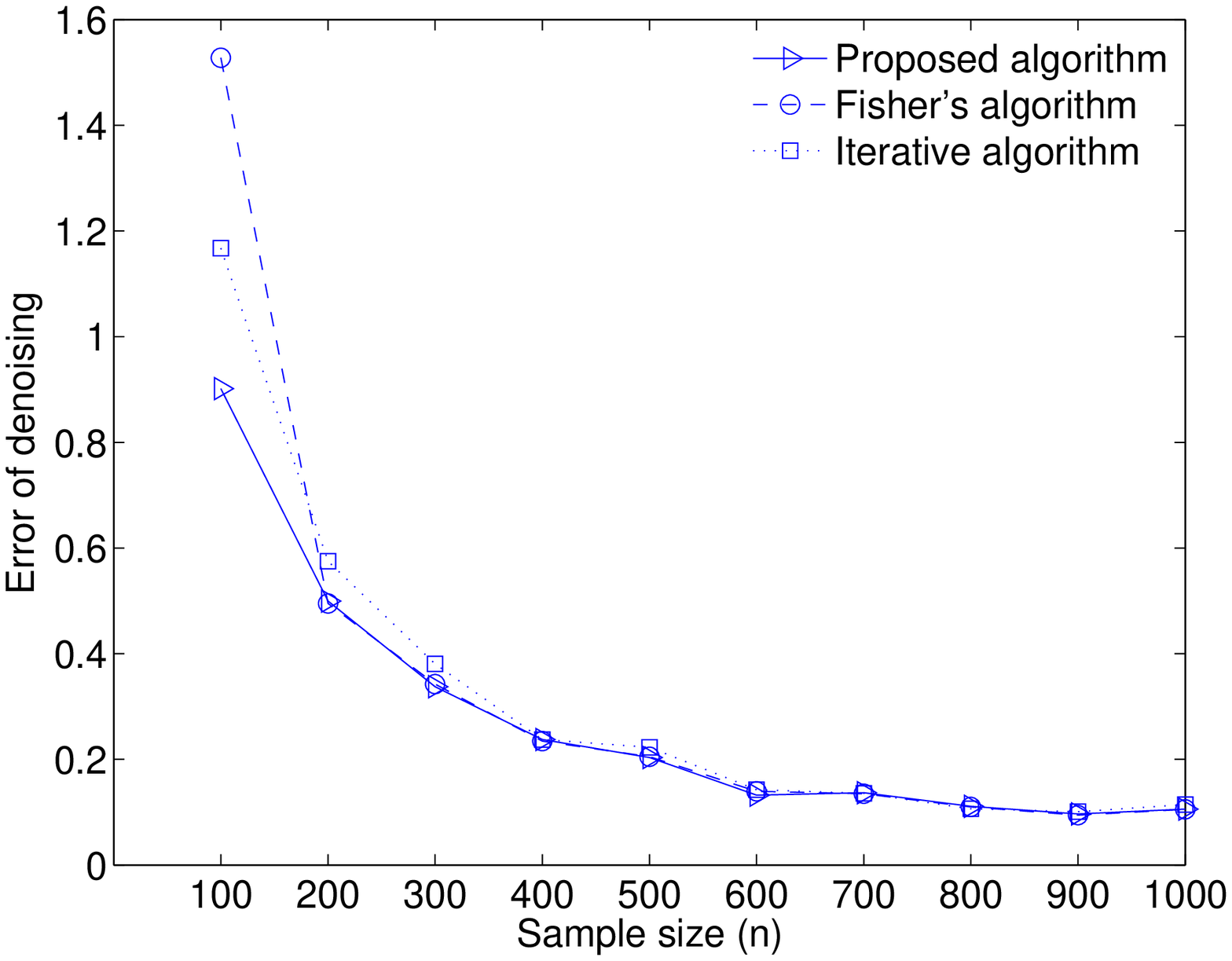}\\
\end{tabular}
\caption{Average misclassification rates (top) and average error of denoising (down) 
in relation to the sample size $n$ obtained with the proposed approach (triangle), Fisher's algorithm (circle) and the iterative version of Fisher's algorithm (square) for the first situation.}
\label{fig. results_sit1}
\end{figure}

\begin{figure}[!h]
\centering
\begin{tabular}{c}
\includegraphics[width=7cm,height=6cm]{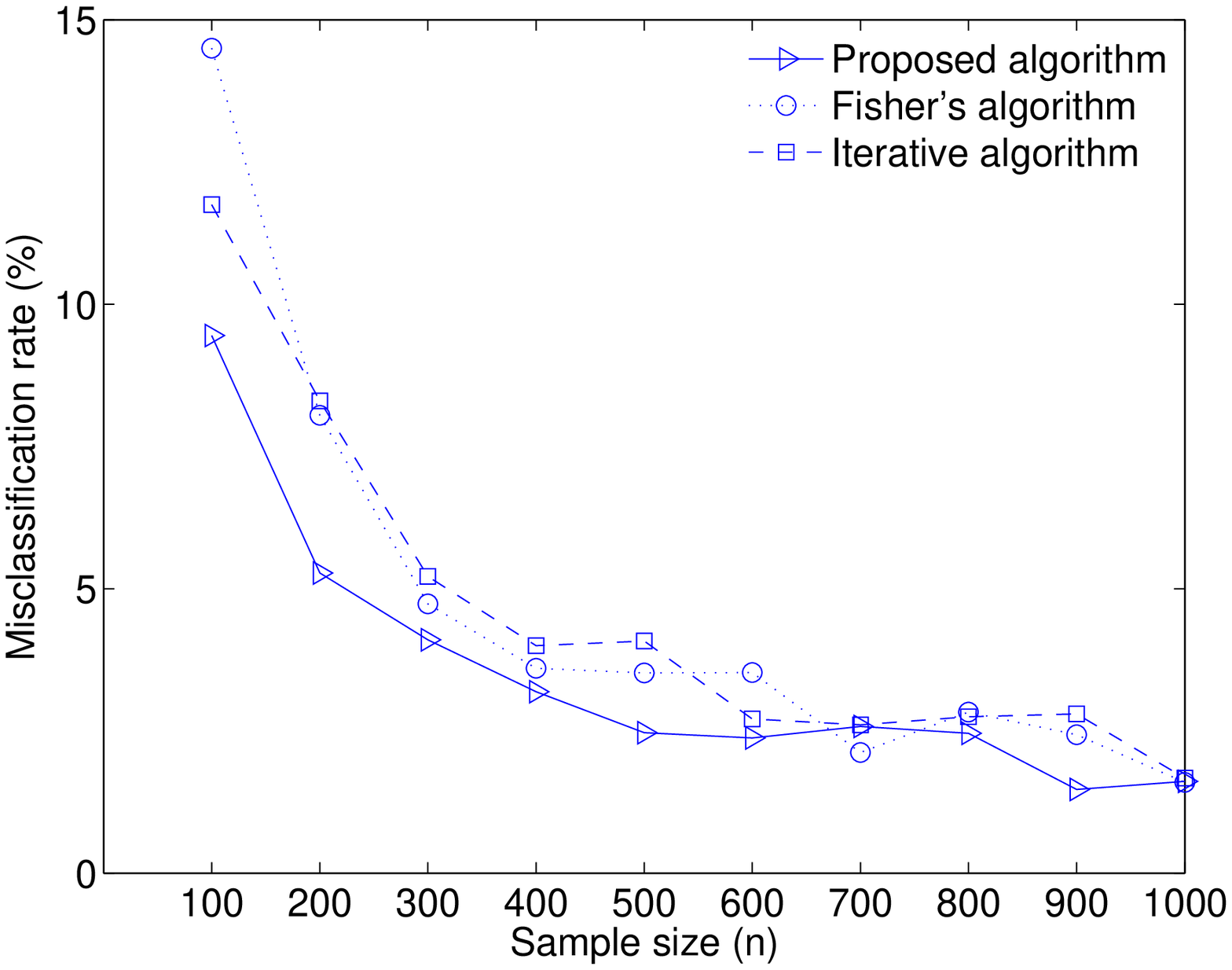}\\
\includegraphics[width=7cm,height=6cm]{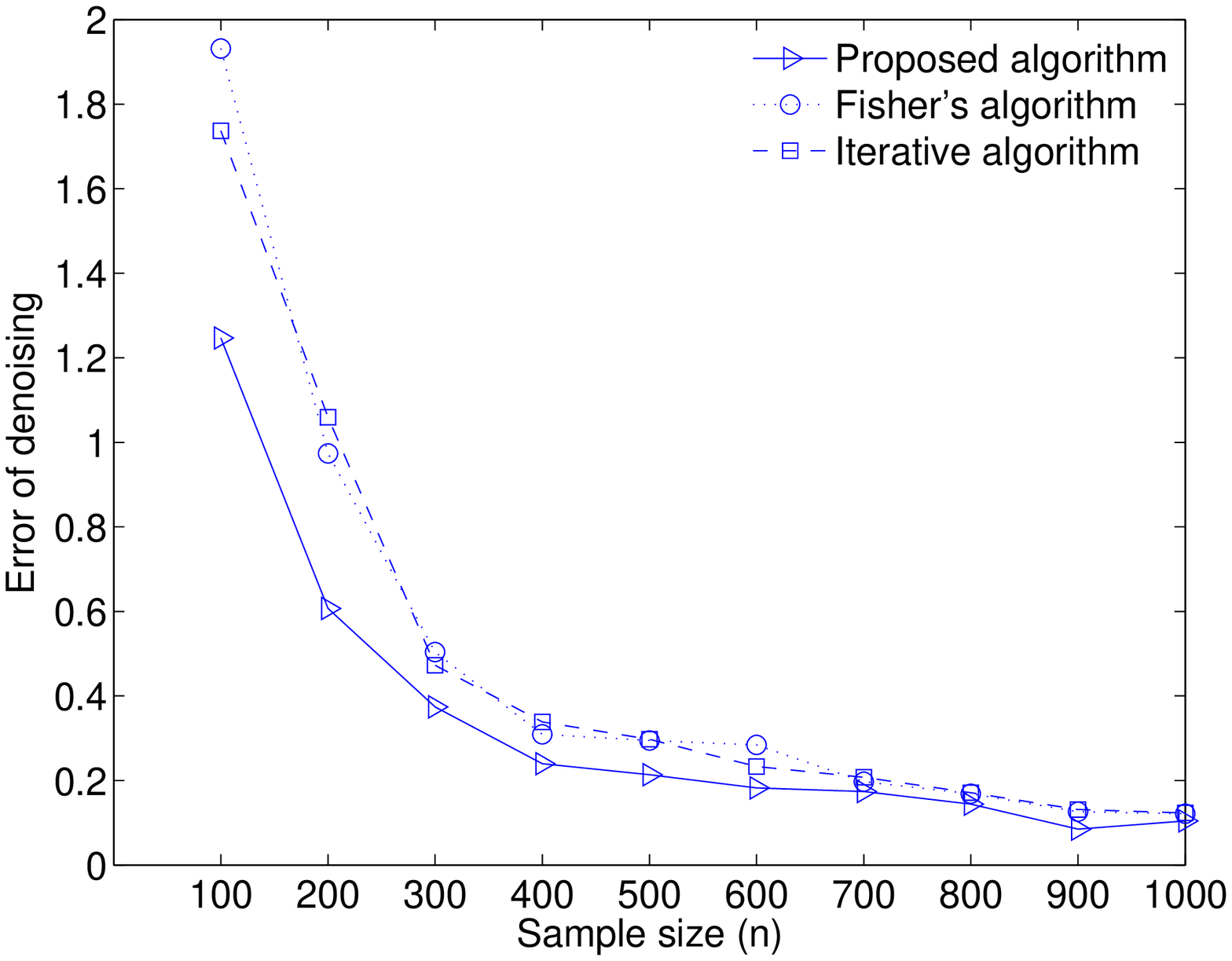}\\
\end{tabular}
\caption{Average misclassification rates (top) and average error of denoising (down) in relation to the sample size $n$ obtained with the proposed approach (triangle),  Fisher's algorithm (circle) and the iterative version of Fisher's algorithm (square) for the second situation.}
\label{fig. results_sit2}
\end{figure}

\begin{figure}[!h]
\centering
\includegraphics[width=7cm,height=6cm]{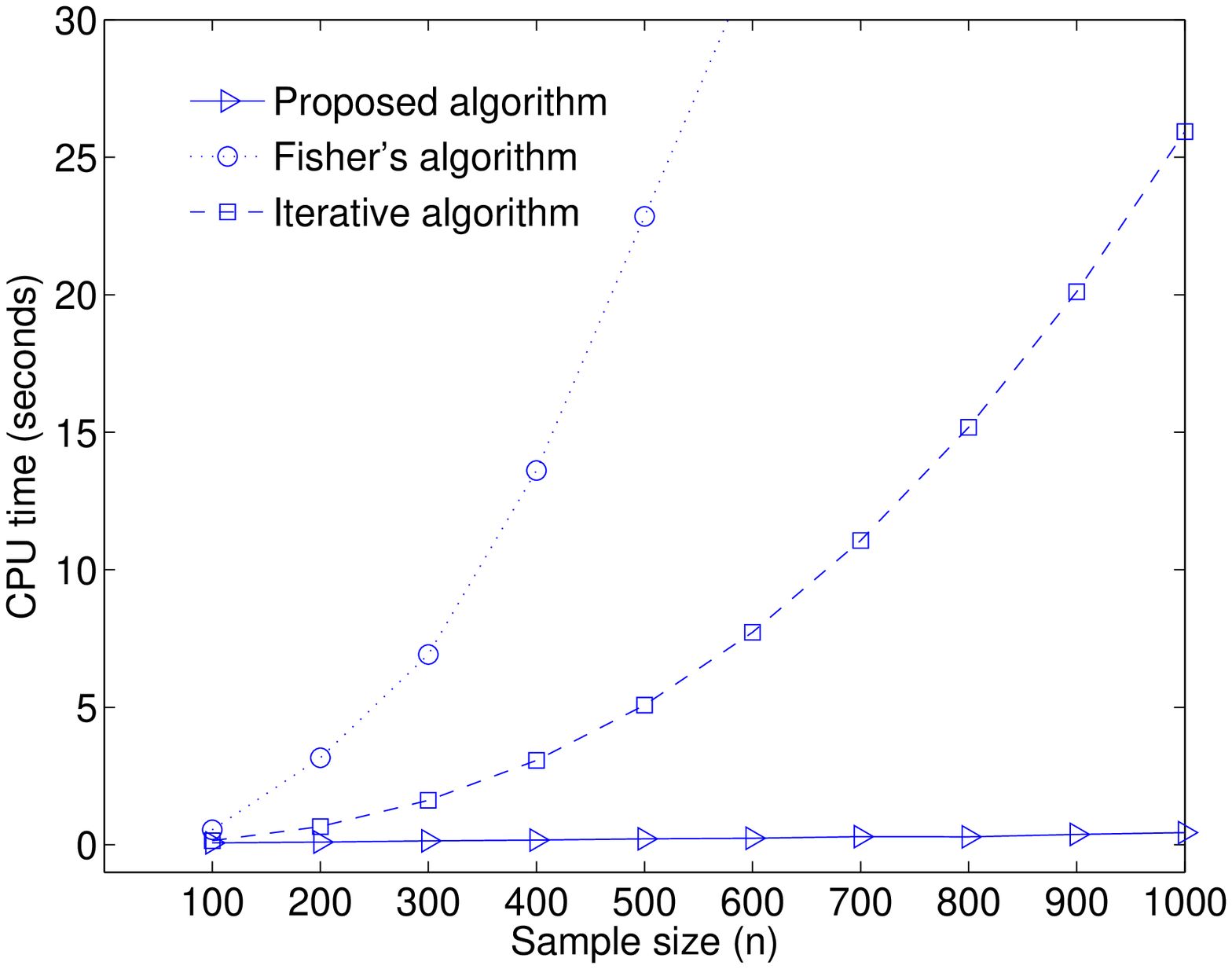} 
\caption{Average running time in relation to the sample size $n$ obtained with the proposed approach (triangle), Fisher's algorithm (circle) and the iterative version of Fisher's algorithm (square).}
\label{fig. cputime}
\end{figure}

Fig. \ref{fig. results_sit1} (top) and Fig. \ref{fig. results_sit2} (top) show the misclassification rate in relation to the sample size $n$ for the two situations of simulated data. It can be observed that the performance of the proposed approach in terms of classification is similar than the global optimization approach. Fig. \ref{fig. results_sit1} (down) and Fig. \ref{fig. results_sit2} (down)  show the error of denoising. The low denoising error obtained by the proposed approach involves a good performance in terms of estimating the true model of the signal, compared to the piecewise regression approaches. Finally, Fig. \ref{fig. cputime} shows the slight variation of the running time of the proposed approach in relation to the sample size. The proposed algorithm is very fast compared to the two other approaches.
\subsection{Real signals}
\label{ssec: exp on real data}

This section presents the results obtained by the proposed approach for signals of switch points operations. Two situations of signals have been considered: one without defect and one with a critical defect. The number of regressive components is chosen in accordance with the number of electromechanical phases of a switch points operation ($K = 5$). The value of $q$ has been set to $1$, which guarantees a segmentation into homogeneous intervals, and the degree of the polynomial regression $p$ has been set to $3$  which is adapted to the different regimes in the signals.

\begin{figure}[!h]
\centering
\includegraphics[width=4.2cm,height=3.4cm]{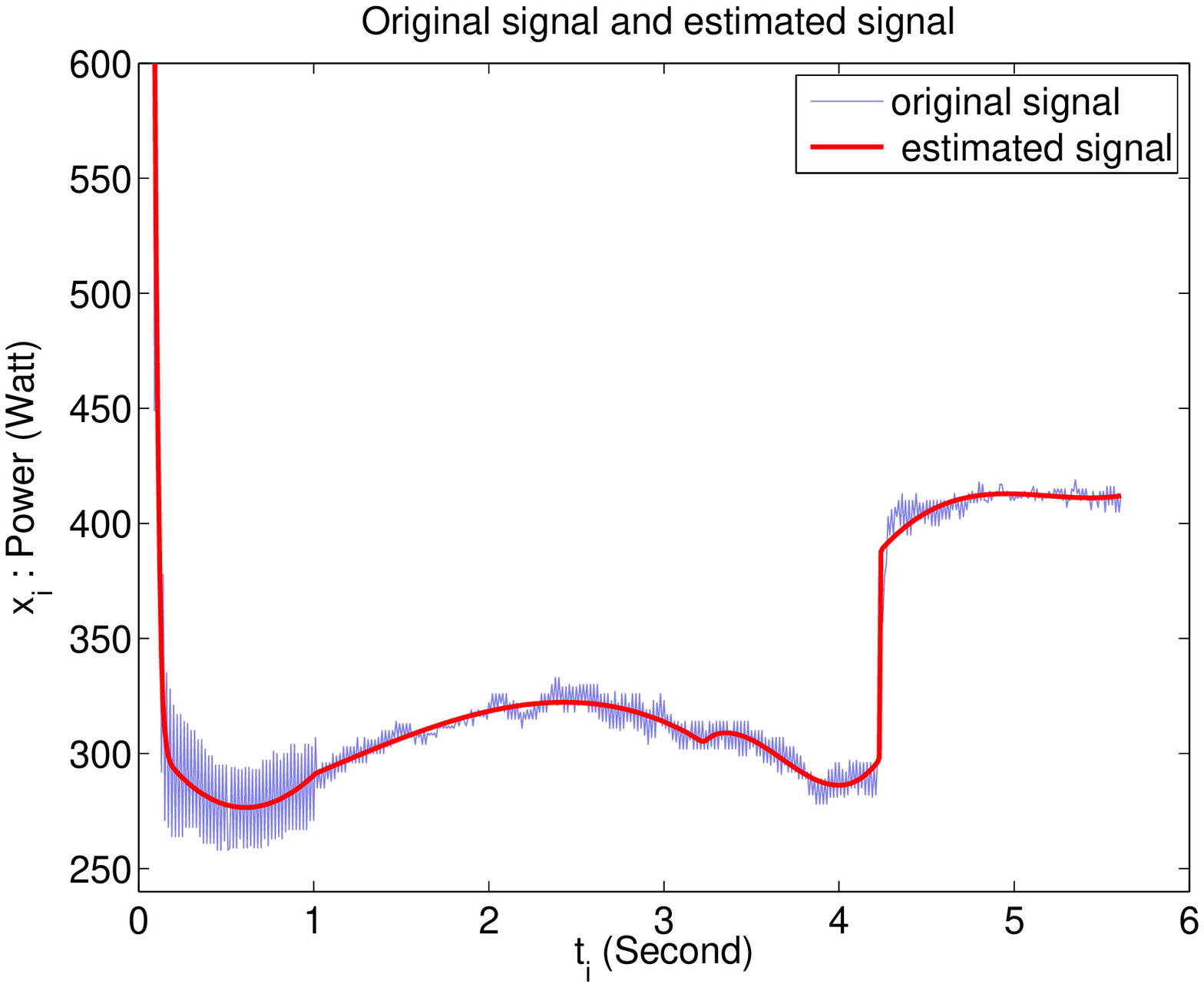}  \includegraphics[width=4.2cm,height=3.4cm]{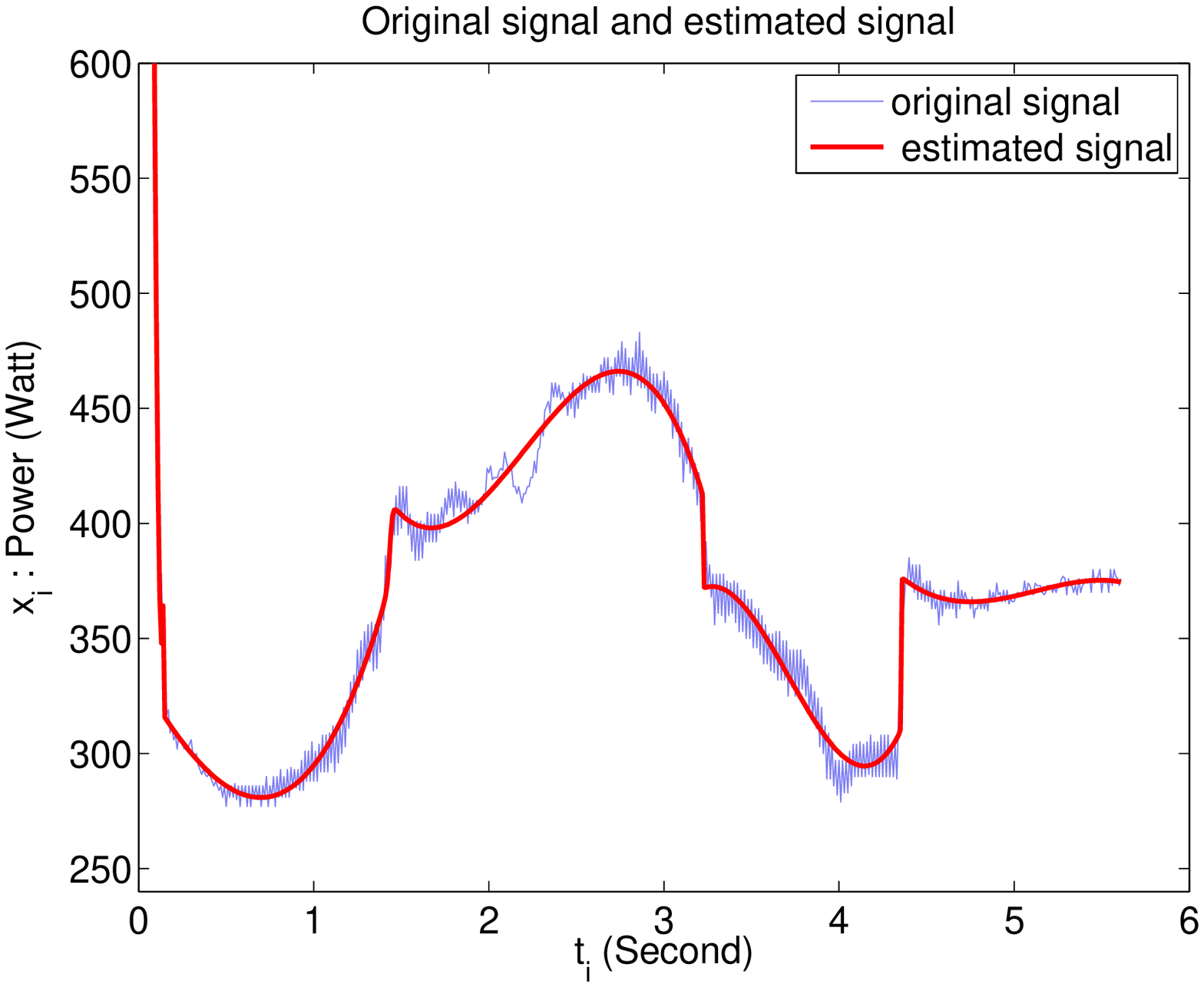} \\
\includegraphics[width=4.2cm,height=3.4cm]{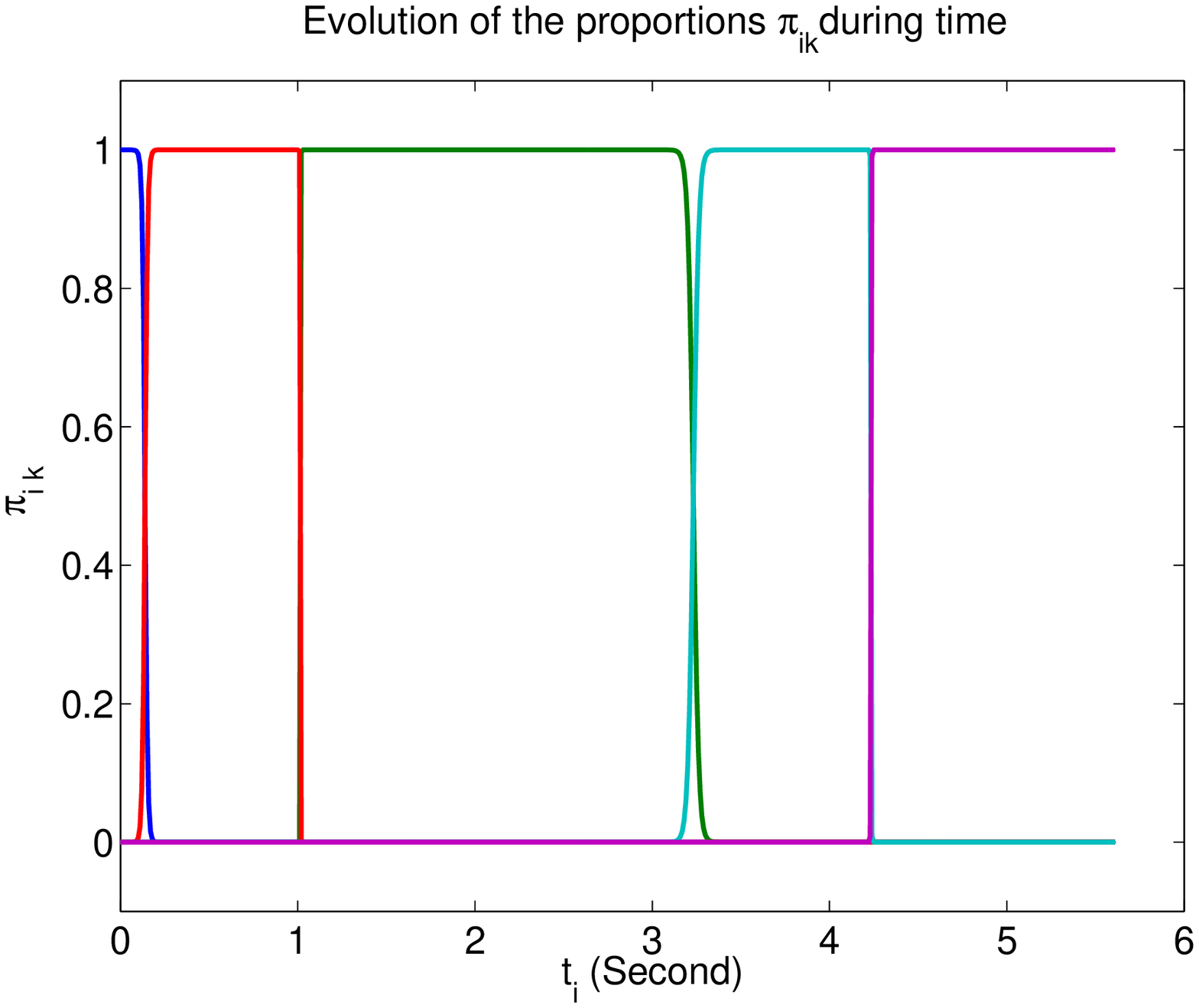}  \includegraphics[width=4.2cm,height=3.4cm]{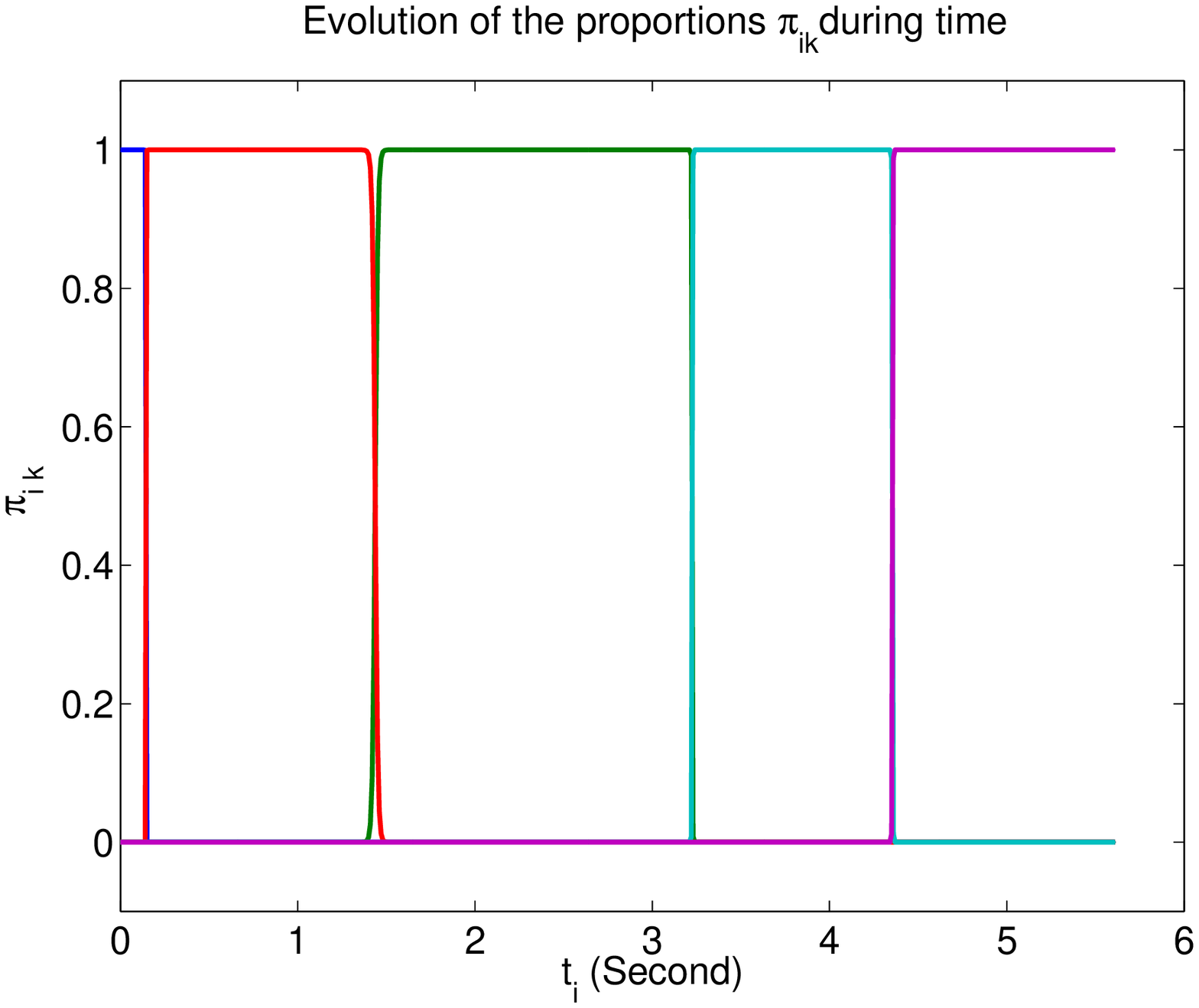}\\
\includegraphics[width=4.2cm,height=3.4cm]{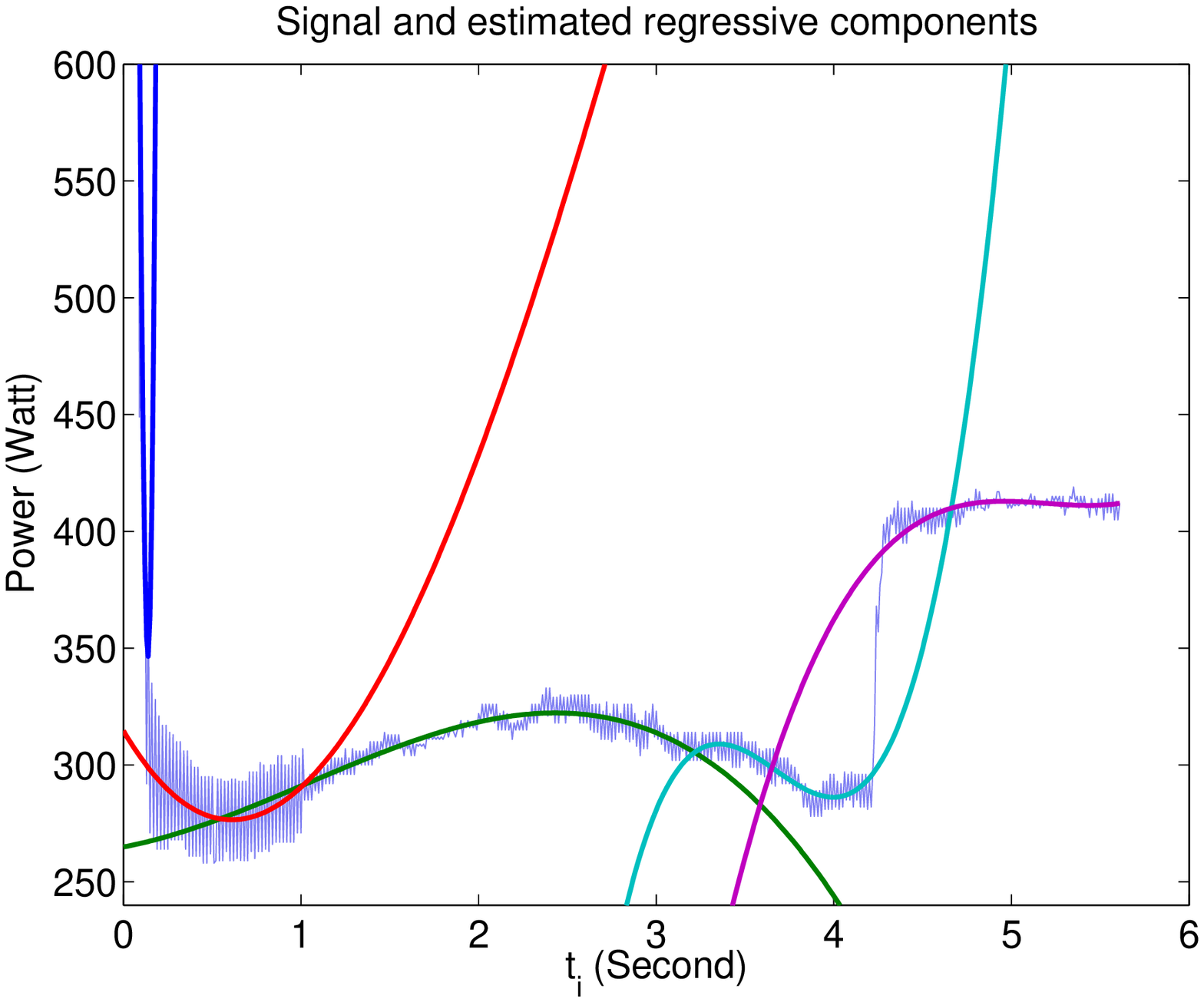}  \includegraphics[width=4.2cm,height=3.4cm]{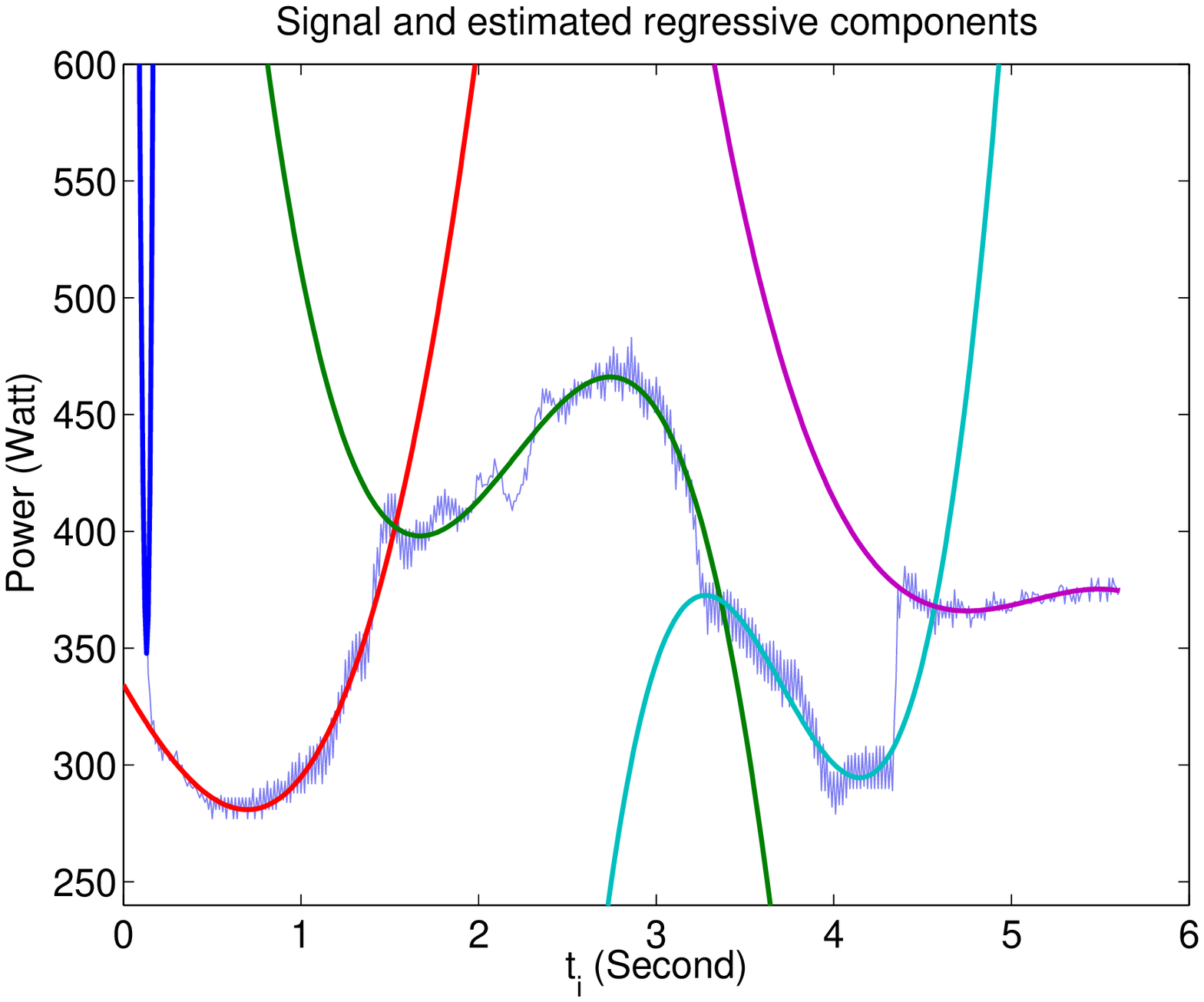}\\
\caption{Obtained results for a signal without defect (left) and for a signal with defect (right).}
\label{resultat_signal_aig_2}
\end{figure}

\begin{figure}[!h]
\centering
\begin{tabular}{cc}
\includegraphics[width=1.55in]{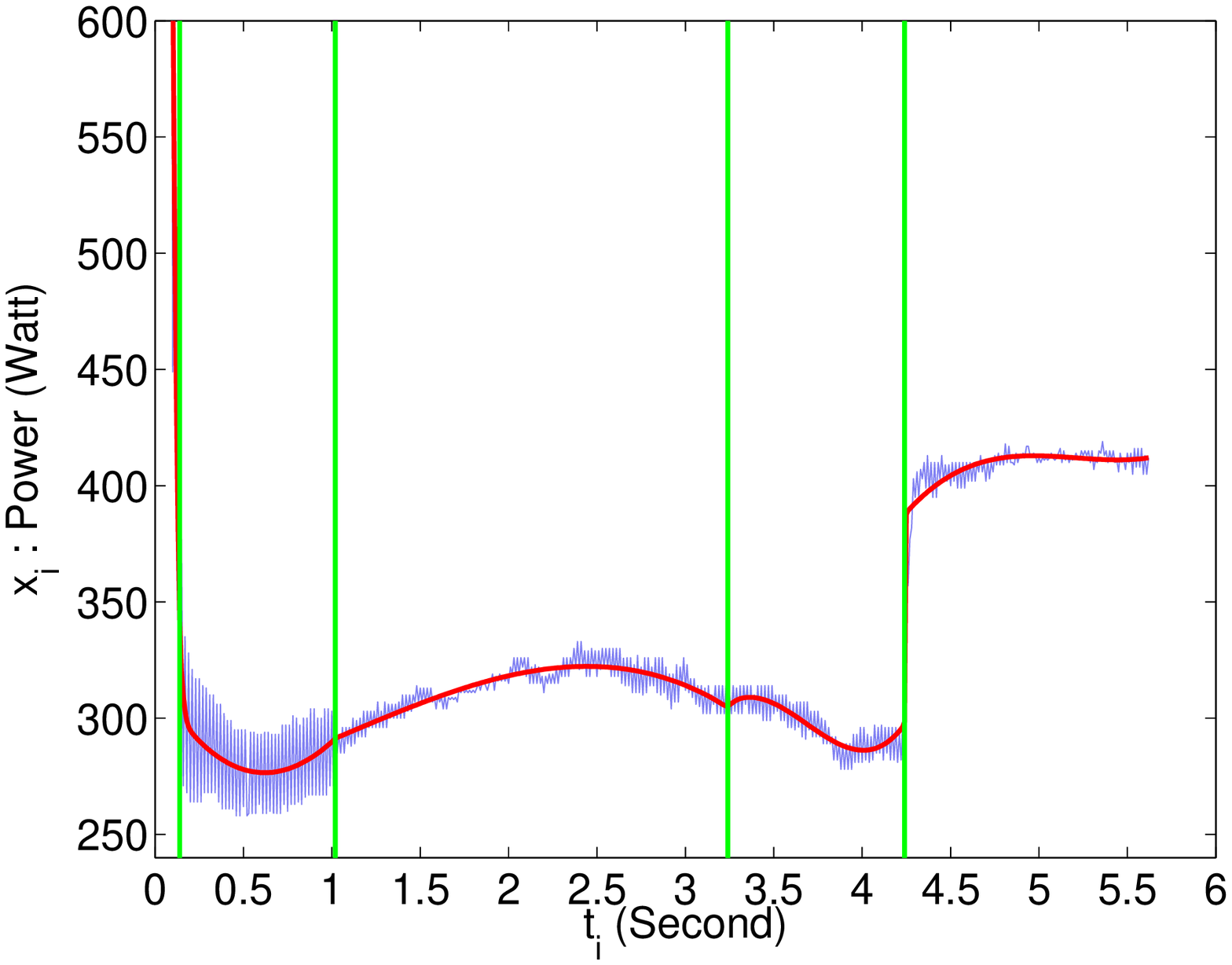} & \includegraphics[width=1.55in]{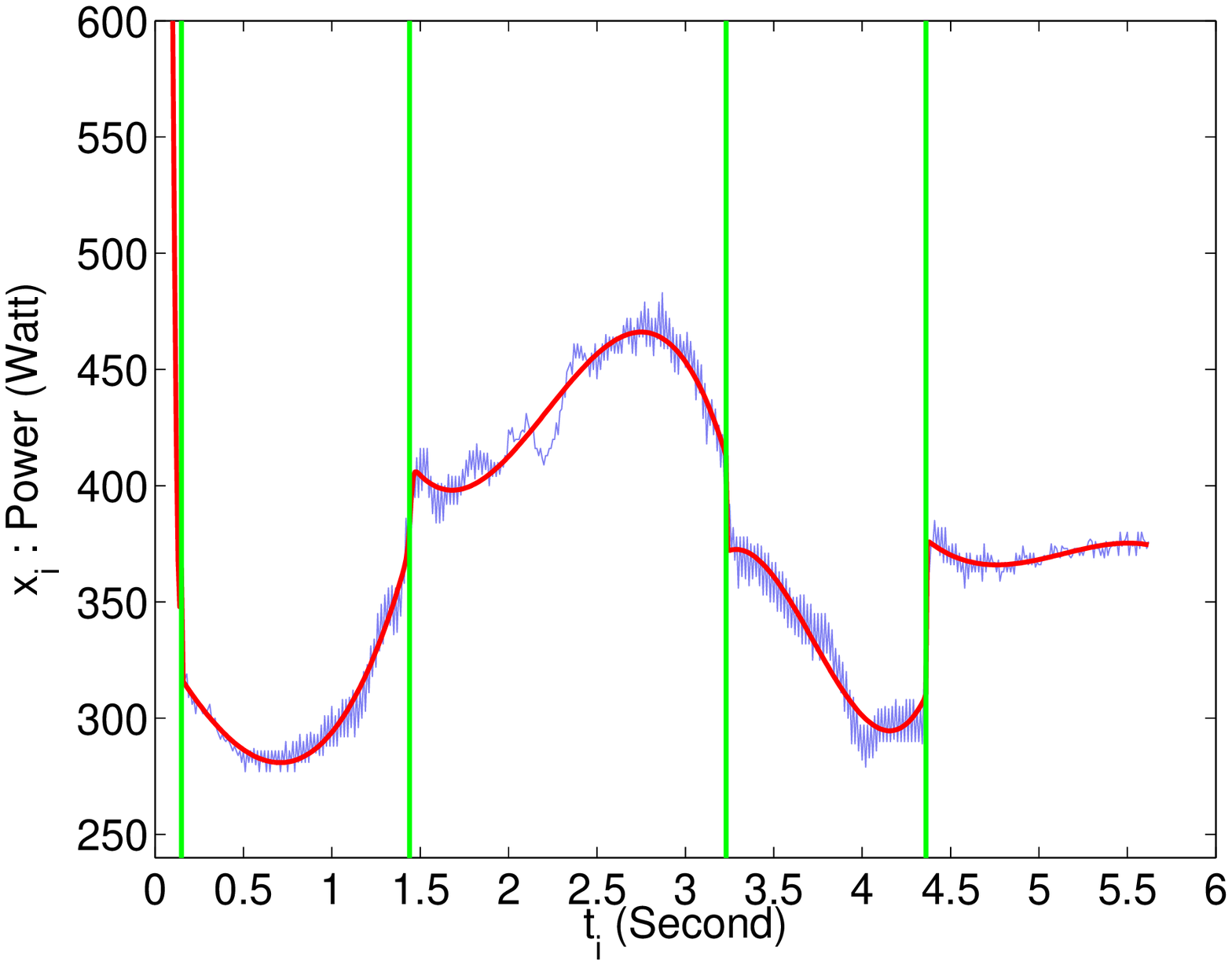} \\
\small{$MSE = 784.932$}& \small{$MSE = 309.789$}\\
\includegraphics[width=1.55in]{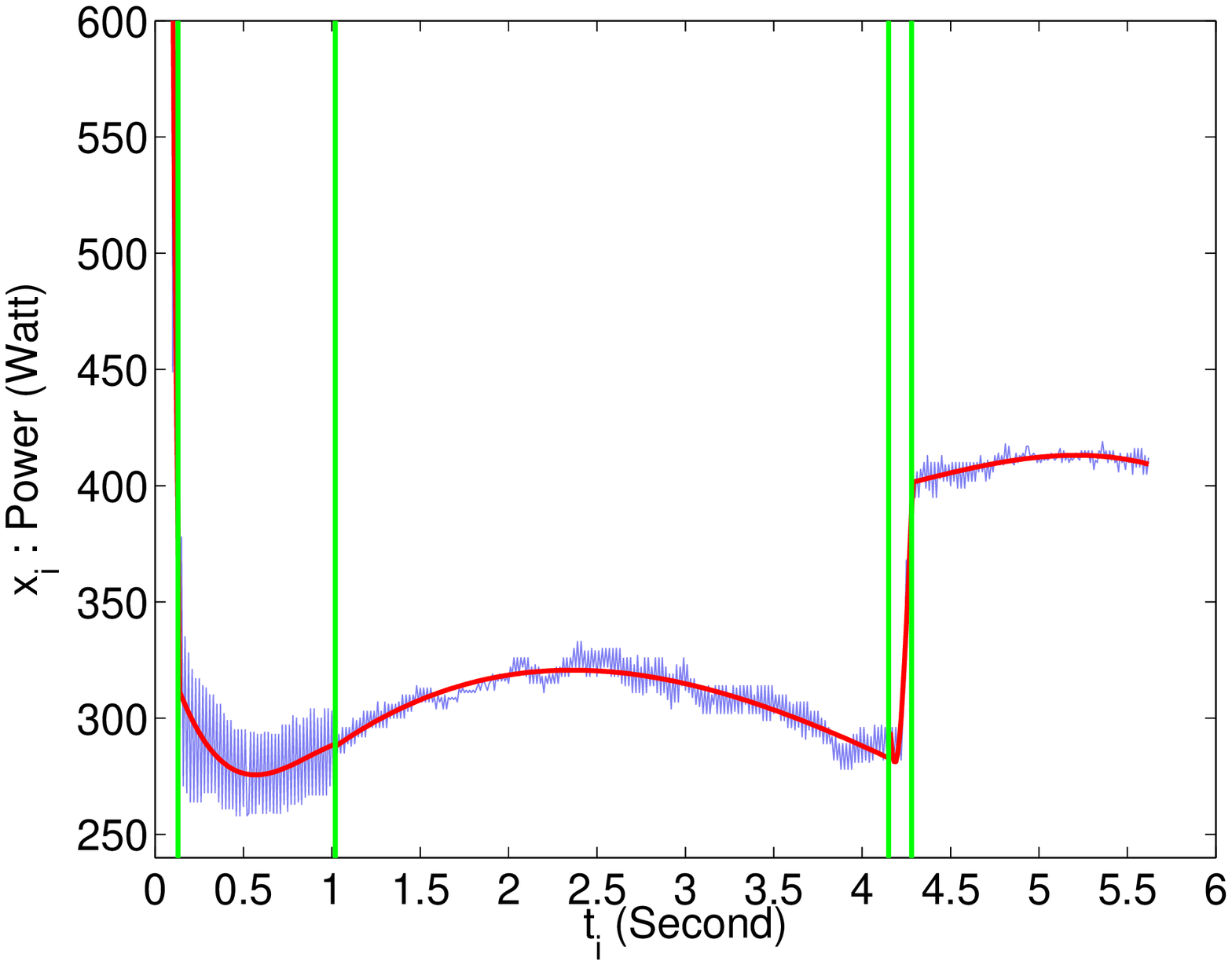} & \includegraphics[width=1.55in]{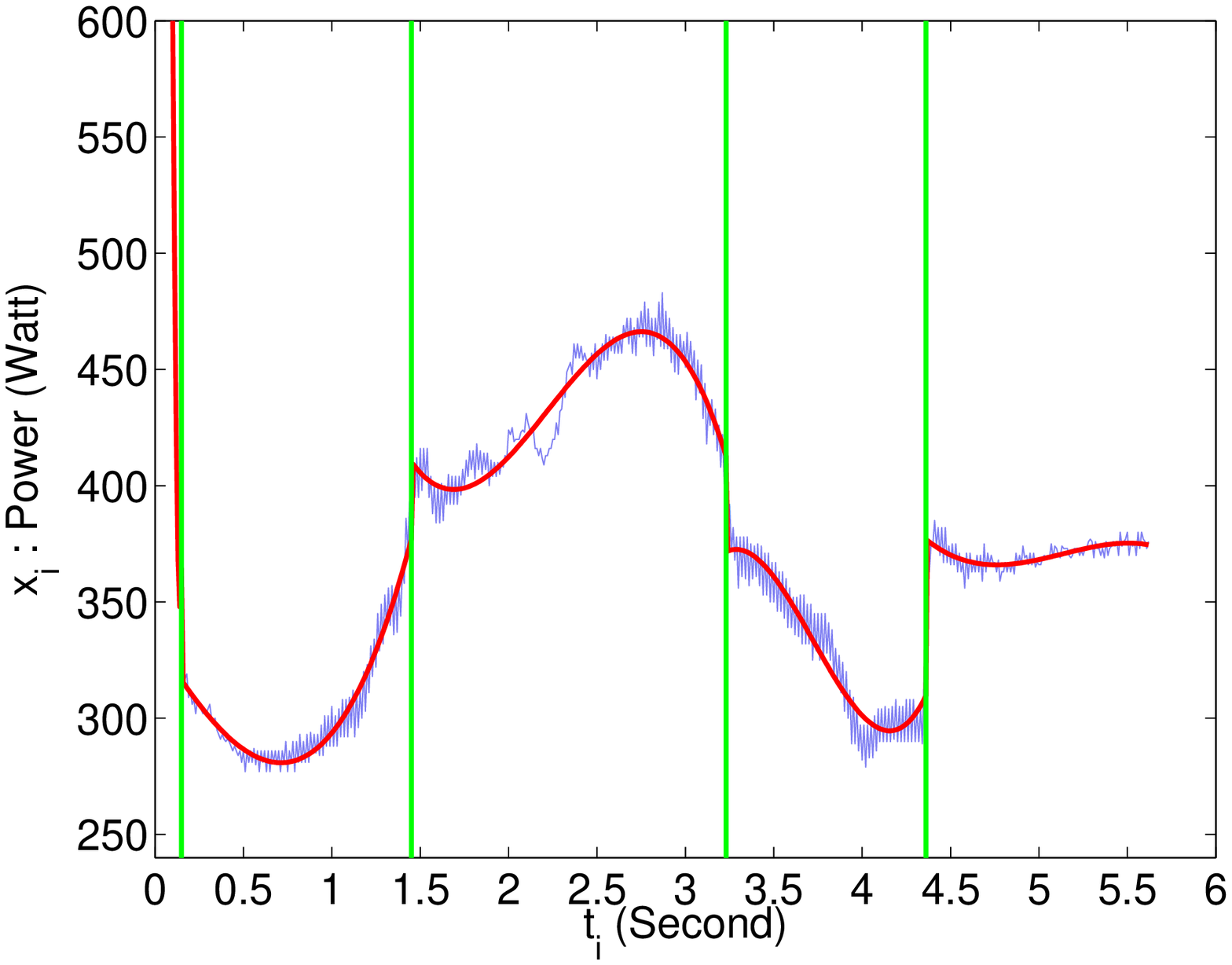}\\
\small{$MSE =  781.639$} & \small{$MSE =  310.251$}\\
\includegraphics[width=1.55in]{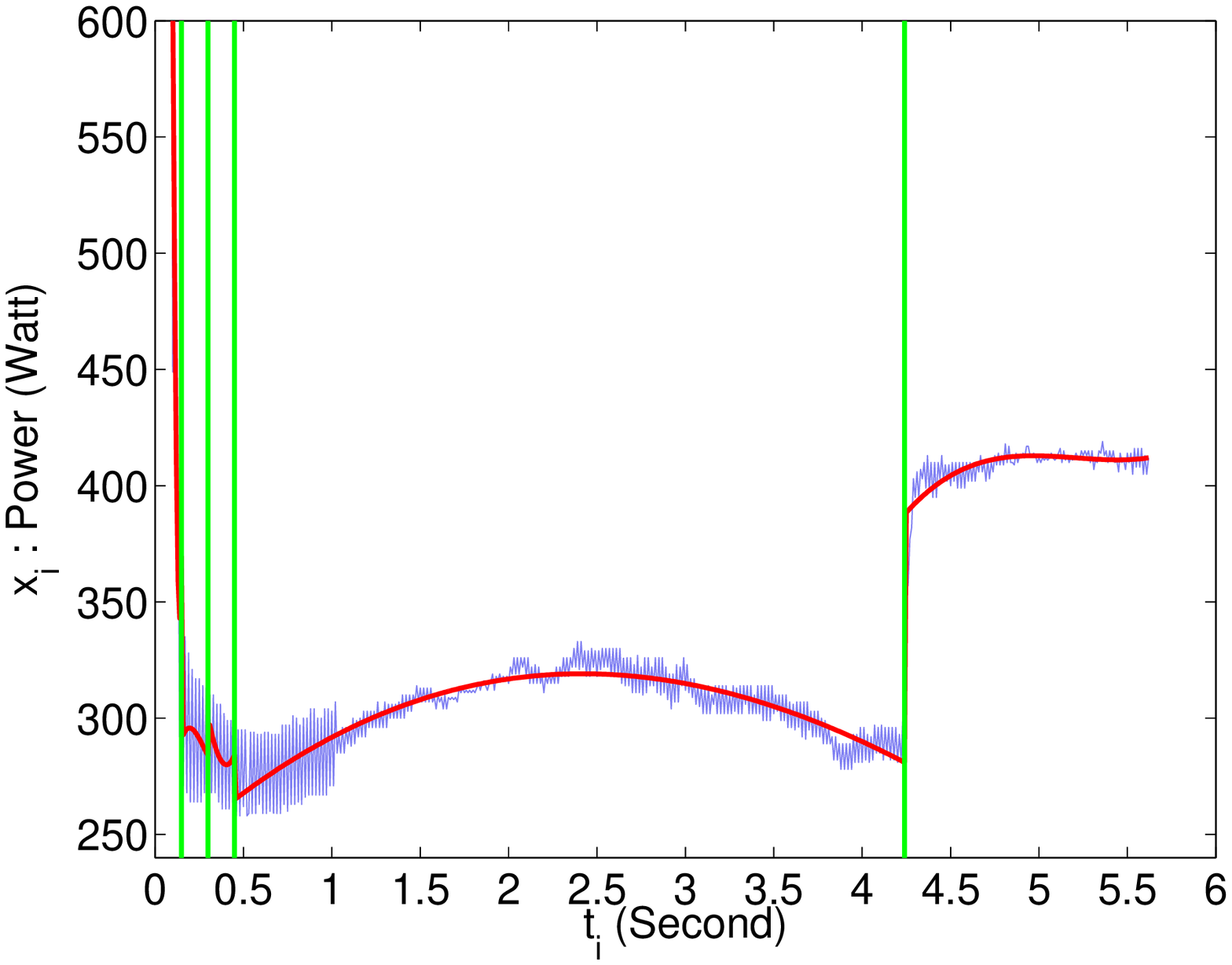}&\includegraphics[width=1.55in]{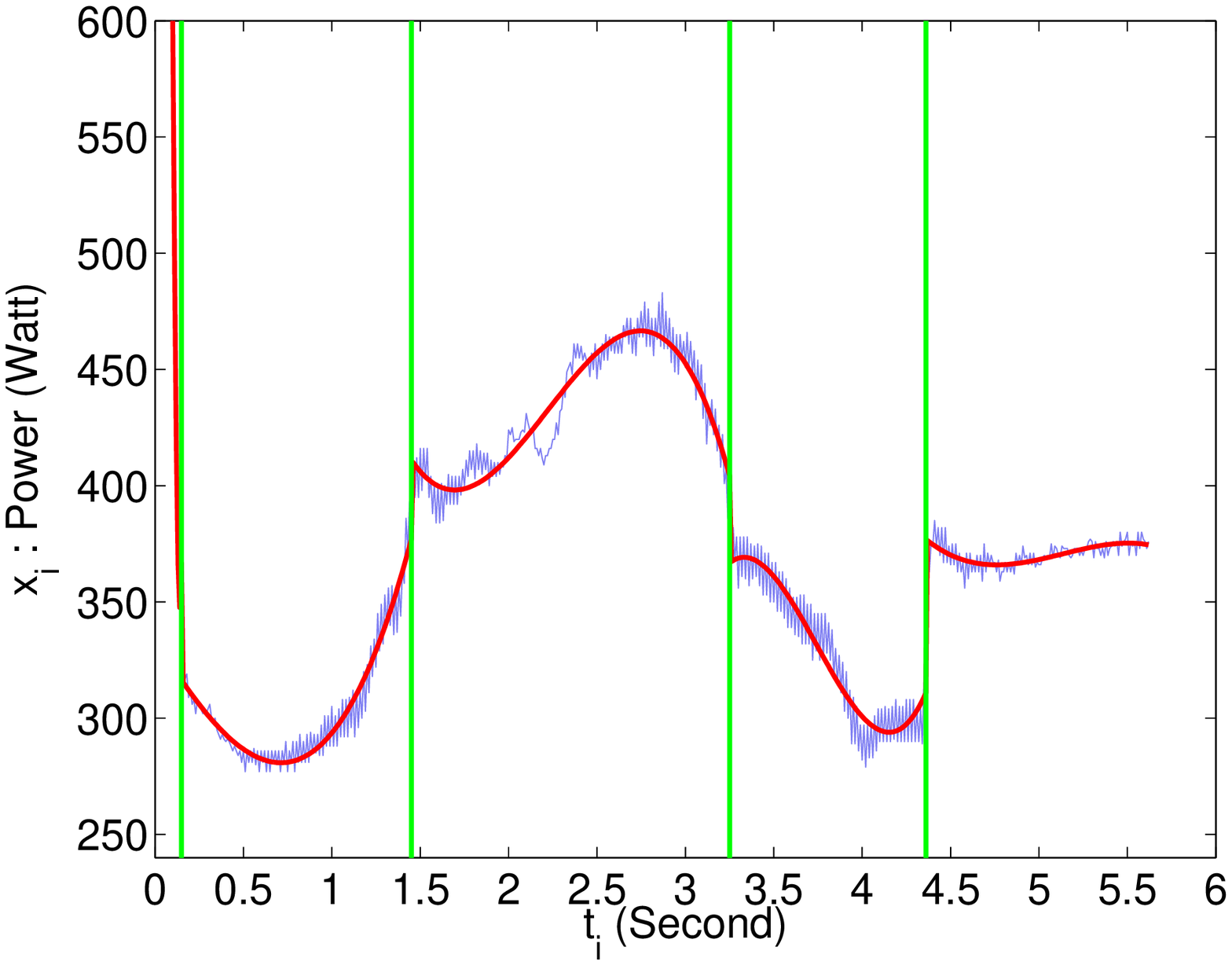} \\
\small{$MSE = 895.558$} & \small{$MSE = 311.001$}
\end{tabular}
\caption{Results obtained by the proposed algorithm (top), Fisher's algorithm (middle) and the iterative version of Fisher's algorithm (bottom) with the estimated model of the signal (in red), the estimated transition points (in green) and the $MSE$ between the original signal and the estimated model.}
\label{fig.resultats_signaux_reels_all}
\end{figure}

Fig. \ref{resultat_signal_aig_2} (top) shows the original signals and the denoised signals (the denoised signal is given by equation (\ref{eq. signal expectation})). Fig. \ref{resultat_signal_aig_2} (middle) shows the variation of the probabilities $\pi_{ik}$ over time. It can be observed that these probabilities are very closed to $1$ when the $k^{th}$ regressive model seems to be the most faithful to the original signal. The five regressive components involved in each signal are shown in Fig. \ref{resultat_signal_aig_2} (down). Fig. \ref{fig.resultats_signaux_reels_all} shows the segmentation, the estimated signals and and the Mean Square Errors ($MSE$) between the original signal and the estimated signal, obtained with the three methods for the two situations of signals. 

\begin{figure}[!h]
\centering
\includegraphics[width=4.2cm,height=3.4cm]{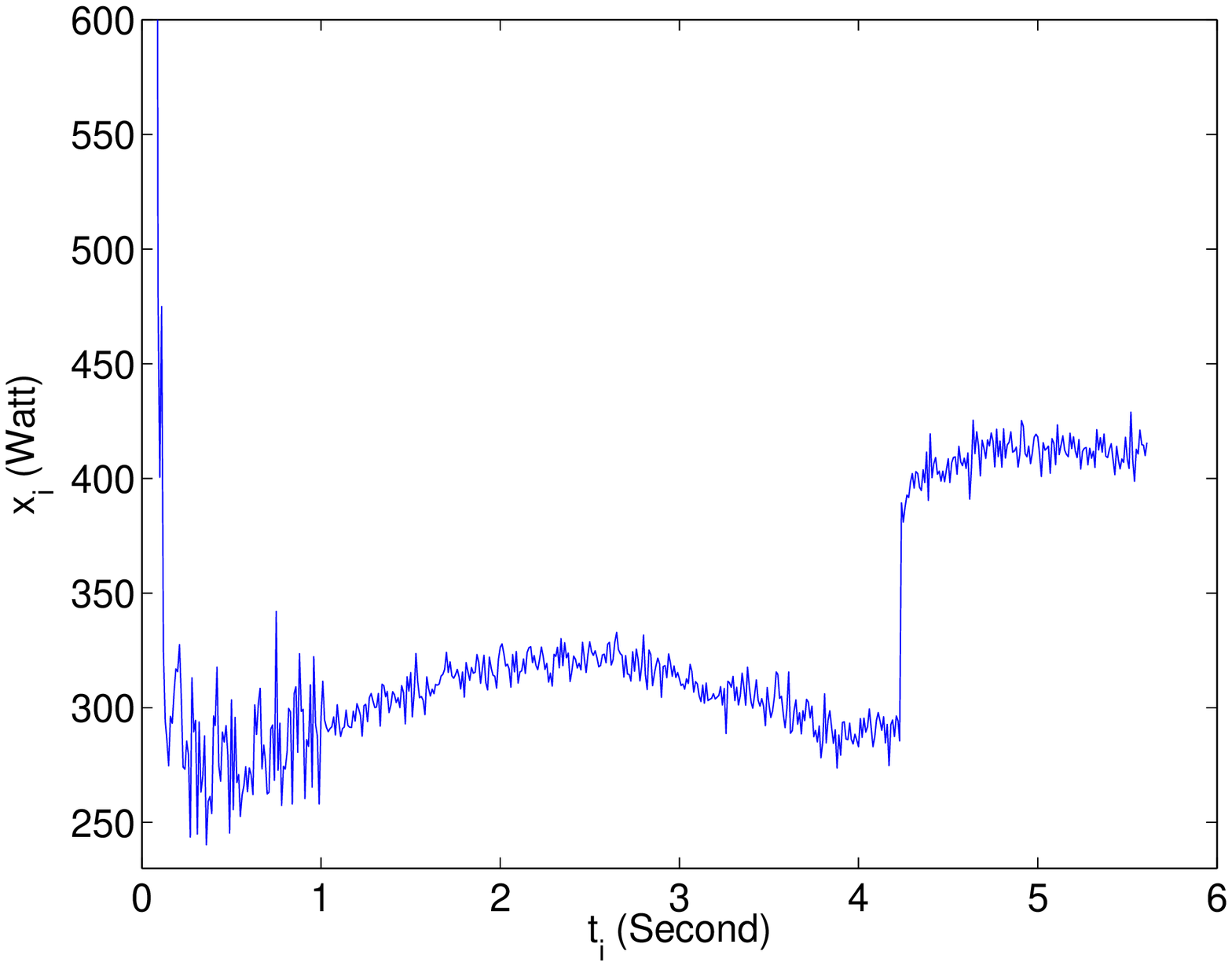}
\includegraphics[width=4.2cm,height=3.4cm]{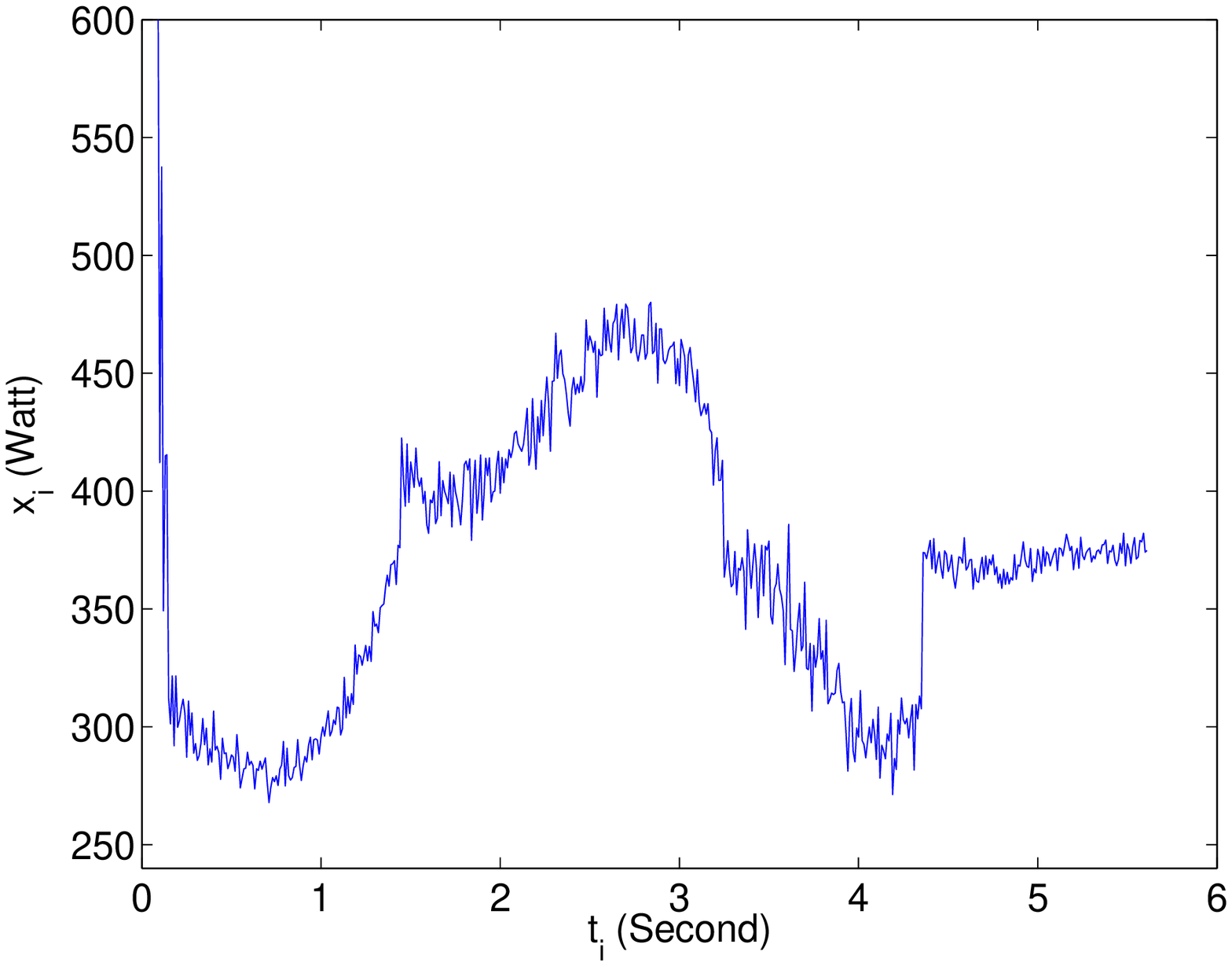}
\caption{Examples of generated signals corresponding to the two considered switch operations.}
\label{genereted signals}
\end{figure}

To illustrate the signal generation model, we generate two signals according to the proposed model using the parameters estimated by the EM algorithm. It can be seen that the generated signals are very similar to the original signals (see  Fig. \ref{genereted signals}).

\section{Conclusion}
\label{sec: conclusion}
In this paper a new approach for feature extraction from time series signals, in the context of the railway switch mechanism monitoring, has been proposed. This approach is based on a regression model incorporating a discrete hidden logistic process. The logistic probability function, used for the hidden variables, allows for smooth or abrupt transitions between polynomial regressive components over time. In addition to signals parametrization, an accurate denoising and segmentation of signals can be derived from the proposed model. The experiments applied to real and simulated data have shown good performances of the proposed approach compared to two algorithms devoted to the piecewise regression.




\begin{thebibliography}{99}
\bibitem{dlr} A. P. Dempster, N. M. Laird and D. B. Rubin, ``Maximum likelihood from incomplete data via the EM algorithm," \emph{Journal of the Royal Statistical Society,} B, 39(1): 1-38, 1977.

\bibitem{bellman} R. Bellman, ``On the approximation of curves by line segments using dynamic programming," \emph{Communications of the Association for Computing Machinery (CACM)} (4), No. 6, pp. 284 June 1961.


\bibitem{fisher} W. D. Fisher, ``On grouping for maximum homogeneity," \emph{Journal of American Statistics. Society} 53, 789-798, 1958.
\bibitem{krishnapuram} B. Krishnapuram, L. Carin, M.A.T. Figueiredo and A.J. Hartemink, ``Sparse multinomial logistic regression: fast algorithms and generalization bounds," \emph{IEEE Transactions on Pattern Analysis and Machine Intelligence,} 27(6): 957-968, June 2005.

\bibitem{mclachlan EM}  G. J. McLachlan and T. Krishnan, \emph{The EM algorithm and extensions,} Wiley series in probability and statistics, New York, 1997.

\bibitem{BIC criterion} G.~Schwarz, ``Estimating the dimension of a model," \emph{Annals of Statistics,} 6: 461-464, 1978.

\bibitem{chen99} K. Chen, L. Xu and H. Chi, ``Improved learning algorithms for Mixture of Experts in multiclass classification," \emph{IEEE Transactions on Neural Networks,} 12(9): 1229-1252, November 1999.

\bibitem{baum welch} L.E. Baum, T. Petrie, G. Soules and N. Weiss, ``A maximization technique occurring in the statistical analysis of probabilistic functions of Markov chains," \emph{Annals of Mathematical Statistics}, 41: 164-171, 1970.

\bibitem{rabiner} L. R. Rabiner, ``A tutorial on hidden Markov models and selected applications in speech recognition," \emph{Proceedings of the IEEE}, 77(2): 257-286, February 1989.

\bibitem{fridman} M. Fridman, Hidden Markov Model Regression, Technical Report, Institute of mathematics, University of Minnesota, December 1993.

\bibitem{jordan HME} M. I. Jordan and R. A. Jacobs, ``Hierarchical Mixtures of Experts and the EM algorithm," \emph{Neural Computation,} 6: 181-214, 1994.

\bibitem{irls} P.~Green, ``Iteratively Reweighted Least Squares for Maximum Likelihood Estimation, and some robust and resistant alternatives," \emph{Journal of the Royal Statistical Society,} B, 46(2): 149-192, 1984.


\bibitem{quandt} R. E. Quandt and and J. B. Ramsey, ``Estimating mixtures of normal distributions and switching regressions," \emph{Journal of the American Statistical Association,} 73, 730-738, 1978.

\bibitem{McGee} V. E. McGee and  W. T. Carleton, ``Piecewise regression," \emph{Journal of the American Statistical Association}, 65, 1109-1124, 1970.


\bibitem{waterhouse} S. R. Waterhouse, \emph{Classification and regression using Mixtures of Experts,} PhD thesis, Department of Engineering, Cambridge University, 1997.









\bibitem{yveslechevalier90} Y. Lechevalier, Optimal clustering on ordered set, Technical report, The French National Institute for Research in Computer Science and Control (INRIA), 1990.

\bibitem{brailovsky} V. L. Brailovsky and Y. Kempner, ``Application of piece-wise regression to detecting internal structure of signal," \emph{Pattern recognition}, 25(11), 1361-1370, November 1992.

\bibitem{ferrari1} G. Ferrari-Trecate and M. Muselli, ``A new learning method for piecewise linear regression," \emph{International Conference on Artificial Neural Networks (ICANN)}, 28-30, Madrid, Spain, August 2002.

\bibitem{sameSFC2007} A. Sam\'{e}, P. Aknin and G. Govaert, ``Classification automatique pour la segmentation des signaux unidimensionnels," Rencontres de la Soci\'{e}t\'{e} Francophone de Classification, ENST, Paris, 2007.
\end{thebibliography}
\end{document}